\documentclass[12pt]{article}
\usepackage{amsmath, amsthm,amsfonts,amssymb}
\usepackage{graphicx}

\usepackage{natbib}

\usepackage{url} 

\newcommand{\blind}{0}

\usepackage{lscape,psfrag}
\usepackage{epsfig,color}
\usepackage{array}
\usepackage{enumerate}
\usepackage{graphics}
\usepackage{epsfig}
\usepackage{verbatim}
\usepackage{cancel}
\usepackage{longtable}
\usepackage{multirow}
\usepackage{float}
\usepackage[normalem]{ulem} 
\usepackage[toc,page]{appendix}
\usepackage{rotating}
\usepackage{amsmath}
\usepackage{url}
\usepackage{bbm}    
\usepackage{amssymb}
\usepackage{amsmath}
\usepackage{graphicx}
\usepackage{hyperref}
\usepackage{natbib}
\usepackage{mathabx}
\usepackage{subfig}
\usepackage{scalerel}
\usepackage{xcolor}
\hypersetup{
	colorlinks,
	linkcolor={purple!80!black},
	citecolor={blue!50!black},
	urlcolor={black!100!black}
}
\usepackage{dsfont}
\usepackage{stmaryrd}
\usepackage{tabularx}
\usepackage{booktabs}

\newcommand {\bs}[1] { {\boldsymbol#1} }

\newcommand {\bmath} {\begin {displaymath} }
\newcommand {\emath} {\end {displaymath} }

\newcommand {\by} {\mathbf{y}}

\newcommand{\bbk}{\boldsymbol{\beta}_k}

\newcommand{\X}{\mathbf{X}}

\newcommand{\xxi}{\mathbf{x}_{i}}
\newcommand{\xxti}{\mathbf{x}_{i}^{\top}}

\newcommand{\xii}{\mathbf{x}_{i}}

\newcommand{\sk}{\sigma^2_k}
\newcommand{\mk}{\boldsymbol{\mu}_k}
\newcommand{\Sk}{\boldsymbol{\Sigma}_k}

\newcommand{\lk}{\lambda_k}

\newcommand{\thita}{\boldsymbol{\theta}_k}
\newcommand{\thitaY}{\boldsymbol{\theta}_k^{\scaleto{Y}{4pt}}}
\newcommand{\thitaX}{\boldsymbol{\theta}_k^{\scaleto{\boldsymbol{X}}{4pt}}}
\newcommand{\ak}{\alpha_k}
\newcommand{\Ok}{\mathbf{\Omega}_k}

\newcommand{\z}{\mathbf{z}}
\newcommand{\Ak}{\mathbf{A}_{k}}
\newcommand{\Tk}{\mathbf{T}_{k}}

\begin{document}

\def\spacingset#1{\renewcommand{\baselinestretch}%
{#1}\small\normalsize} \spacingset{1}


\if0\blind
{
  \title{\bf Bayesian Cluster Weighted Gaussian Models}
  \author{Panagiotis Papastamoulis\thanks{{\tt papastamoulis@aueb.gr}}\hspace{.2cm}\\
    Department of Statistics, Athens University of Economics and Business, Greece\\
    and \\
    Konstantinos Perrakis \\
        Department of Statistics, Athens University of Economics and Business, Greece}
  \maketitle
} \fi

\if1\blind
{
  \bigskip
  \bigskip
  \bigskip
  \begin{center}
    {\LARGE\bf Bayesian Cluster Weighted Gaussian Models}
\end{center}
  \medskip
} \fi

\bigskip
\begin{abstract}
We introduce a new framework for Bayesian inference in mixtures of normal linear regression models with random covariates. Such types of mixtures belong to the category of cluster-weighted models. 
The proposed Bayesian cluster-weighted model aims to encompass potential heterogeneity in the distribution of the response variable as well as in the multivariate distribution of the covariates for detecting signals relevant to the underlying latent structure. Of particular interest are potential signals originating from: (i) the linear predictor structures of the regression models and (ii) the covariance structures of the covariates. We model these two components using a lasso shrinkage prior for the regression coefficients and a graphical-lasso shrinkage prior for the covariance matrices. A fully Bayesian approach is followed for estimating the number of clusters, by treating the number of mixture components as random and implementing a trans-dimensional telescoping sampler. Alternative Bayesian approaches  based on overfitting mixture models or using information criteria to select the number of components are also considered. The proposed methodology is compared to mixtures of regressions, mixtures of experts and existing cluster-weighted models in simulation studies and an application to a biomedical dataset.
\end{abstract}

\noindent%
\vfill

\newpage
\spacingset{1.75} 

\section{Introduction}
\label{intro}

In many applications of  regression analysis, data arise from populations that are not homogeneous but instead consist of unobserved subgroups with different patterns of association between covariates and responses. Mixtures of regressions provide a way to capture this type of latent heterogeneity, so that each cluster is characterized by its own regression relationship. Typically, in a regression setting, the covariates are considered fixed; however, this is not always a realistic assumption. Random covariates can occur in any application where it is not possible to control their values. A well-known example can be found in \cite{Hosmer} dealing with halibut growth, where the relationship between length and age differs between males and females, and as the fish originate from commercial catches, it is not possible to control for age \citep{hennig2000identifiablity}, leading to variability in the covariates which has to be taken into account. Situations of this kind are common in practice: both the response–covariate relationship and the distribution of the covariates may vary across subpopulations. Models that allow for mixtures of regressions with random covariates, such as cluster-weighted models \citep{gershenfeld1997nonlinear}, are therefore a natural and useful framework for describing such data and for uncovering meaningful group structure. According to \cite{punzo2015parsimonious}, the main benefit when considering random covariates in a mixture of regressions is that they affect the clustering structure, a property which is known as assignment dependence \citep{hennig2000identifiablity}.

To put matters into context, let us consider observed data consisting of independent pairs $\{(y_i, \mathbf x_{i}^\top)^\top, i = 1,\ldots,n\}$, where $n$ denotes the sample size. Each observation  consists of a continuous response variable (denoted as $y_i$) which could be linearly affected by a subset of $p$  covariates, denoted as $\mathbf x_i = (x_{i1},\ldots,x_{ip})^\top$, for  $i=1,\ldots,n$. We consider cases where the population exhibits heterogeneity, i.e., it consists of an unknown number of latent groups (or clusters), where each cluster exhibits distinct patterns in the underlying (linear) relation of the response ($y$) to the covariates ($\mathbf x$).  The goal then, is to infer the number of hidden clusters as well as their underlying distributional characteristics. Under a model-based clustering point of view \citep{banfield1993model, bensmail1997inference, fraley2002model, mcnicholas2016model, grun2019model},  the focus is on $p(y| \mathbf x)$ (the conditional distribution of the response given the covariates) and, typically, each cluster is  associated with a mixture component (note, however, that the reverse correspondence does not necessarily hold; 
see discussion in Section \ref{sec:K}). The nature of the modelling approach then depends on the  assumption imposed on $\mathbf x$, which may lead to mixtures of regressions and mixtures of experts when the covariates are considered to be fixed or, alternatively, to cluster-weighted mixtures with random covariates. Below we provide a brief overview of these types of models and elaborate upon our contribution.

\textbf{Mixtures of regressions.} Under this modelling approach
the covariate matrix is considered fixed and, in addition, it does not affect the clustering structure, i.e., the assignment of data points to latent groups is not influenced by $\mathbf x$ directly. Within a maximum likelihood framework,  estimation is commonly performed using the expectation-maximization (EM) algorithm \citep{Dempster_etal77} or variants \citep{mclachlan2008algorithm}.
The related literature is vast, starting from mixtures of linear regression models \citep[see, e.g.,][]{desarbo1988maximum} and extending to mixtures of generalized linear models \citep{jansen1993maximum, wedel1995mixture, gaffney1999trajectory}, robust alternatives based on the $t$-distribution \citep[see, e.g.,][]{yao2014robust} and penalized approaches \citep{reqularizedFMR, khalili_lin2013, Städler2010, chamroukhi2016unsupervised}, among others. Other models include semi-parametric  \citep{hunter2012semiparametric, huang2012mixture, berrettini2023semiparametric}  or non-parametric \citep{huang2013nonparametric} mixtures of regressions. In general, implementation of such models is well established; see, e.g., \cite{flexmix1, flexmix2, flexmix3}. 
It is worth noting that Bayesian approaches within this modelling framework are also abundant; see, for instance,  \cite{viele2002modeling, hurn2003estimating, shi2005hierarchical, lee2016bayesian, papastamoulis2023model, im2023bayesian, cozzini2014bayesian} among other works.  
Regardless of the estimation procedure, the main characteristic of mixtures of regressions  is that group assignment is solely based upon the conditional distribution of the response variable; as a result,  any potential discriminant signal about the latent group structure in $\mathbf x$ is not taken into account. 

\textbf{Mixtures of experts.} This category consists of a broader class of mixture models where mixing probabilities are modelled as functions of the covariates (which are still considered fixed); the function linking the probabilities to the covariates defines the so-called ``gating network'' and the part of the model that predicts $y$ from $\mathbf x$ (the conditional distribution of the response variable) is the ``expert network'' \citep[see, e.g.,][]{https://doi.org/10.1002/widm.1246, gormley2019mixture}. The origins of such models can be traced  in the machine learning community  \citep{jacobs1991adaptive}, with various applications found in speech recognition \citep{peng1996bayesian}, election studies \citep{10.1214/08-AOAS178}, social networks \citep{gormley2010mixture} and time-series data analysis \citep{fruhwirth2012labor}, among others.  \cite{nguyen2016laplace} and \cite{CHAMROUKHI201620} present mixtures of experts with Laplace and $t$ distributed errors, respectively. \cite{murphy_murphy2020} introduced a parsimonious family of Gaussian mixtures of experts, which is also available as an \textsf{R} \citep{R} package \citep{moeclust}. A general review, discussing mixtures of regression models and mixtures of experts, is provided in \cite{wedel2002concomitant}. 

\textbf{Cluster weighted models.} 
These models (abbreviated accordingly as CWMs), which treat covariates as random, were initially introduced in \cite{gershenfeld1997nonlinear} \citep[see also][]{Gershenfeld1999}. They were later formalized in the works of \cite{ingrassia2012local}, \cite{ingrassia2014modelbased}, and \cite{ingrassia2015erratum, ingrassia2015}, providing extensions involving elliptical distributions, robust alternatives such as the use of the $t$-distribution, and generalized linear models, respectively.
There exist numerous extensions in the related literature for CWMs, including  models that incorporate skewed distributions \citep{gallaugher2022multivariate}, multivariate responses \citep{dang_etal2017}, models for factor analysis \citep{subedi2013clustering} and models with parsimonious parameterizations as in \cite{punzo2015parsimonious}. Further extensions include matrix-variate data analysis as considered in \cite{tomarchio2021matrix} and \cite{gallaugher2022matrix}, models for functional data \citep{Anton2025cluster} and parsimonious models with further modelling layers dealing with potential heterogeneity within clusters \citep{oh2023merging}.
In other works, the CWM idea is incorporated within the framework of hidden Markov models for longitudinal and time-series data analysis \citep[see, e.g.,][]{punzo_ETAL_2021,tomarchio_etal_2024}.
Implementation of CWMs is generally well-established under the frequentist framework through the use of  EM algorithms and variants; see, e.g., \cite{flexcwm}. Bayesian approaches within this modelling framework are limited. \cite{perrakis2023regularized} proposed a joint mixture model with random covariates under a semi-Bayesian approach, using shrinkage priors for sparse estimation of regression coefficients, considering Laplace \citep{park_casella} and normal-Jefrreys \citep{Figueiredo2001} priors, and also of the covariance matrices of the predictors via the graphical-lasso prior \citep{Wang2012}.
It should be noted that the methodological approach in \cite{perrakis2023regularized} is not fully Bayesian as it is: (i) restricted to point estimation via expectation-conditional maximization  (ECM) algorithms \citep{Meng_Rubind1993}, (ii) dependent on tuning procedures for penalty parameters and (iii) designed for models with a fixed number of components where potential determination of clusters relies on information criteria. 

In this work we present a new Bayesian Gaussian CWM (BGCWM) approach. The proposed model allows direct uncertainty quantification for all related aspects of latent structure. Specifically, the proposed model treats the number of mixture components as unknown by using the flexible framework of generalized mixtures of finite mixture models of \cite{fruhwirth2021generalized} (see \citealp{10.1214/08-BA304, miller2018mixture} for additional information). Furthermore, the model  incorporates shrinkage of (i) the regression coefficients via the Bayesian lasso prior \citep{park_casella} and (ii) the covariance structure of the covariates via the Bayesian graphical lasso \citep{Wang2012}, using half-Cauchy hyperpriors \citep{Polson_Scott_2012} for the corresponding penalties, thus, avoiding potential tuning problems.  Markov chain Monte Carlo (MCMC) sampling is based on an augmented Gibbs algorithm where all the full conditional distributions of the parameters are available and only one additional Metropolis-Hastings step for a single parameter is required when the number of clusters is unknown. 

The rest of the paper is organised as follows. In Section \ref{sec:model} we provide the general description of the model; defining initially the likelihood function of the model in Section \ref{sec:likelihood} and continuing with a description of the prior distributional assumptions related to the mixture components, the number of clusters and the related parameters of the linear regressions and of the covariates in Section \ref{sec:prior}. Posterior inference based on MCMC sampling is discussed in Section \ref{sec:MCMC}; the full conditional distributions (conditional on the number of mixture components) are described in Section \ref{sec:gibbs} and  the implementation of telescoping sampling 
for updating the number of mixture components is presented in Section \ref{sec:telescoping}. Identifiability issues and post-hoc variable evaluation are discussed in Sections \ref{sec:identifiability} and \ref{sec:var_sel}, respectively. In Section \ref{sec:sim} we consider simulation studies and comparisons to other methods, while an application to a real dataset is presented in Section \ref{sec:real}. A synopsis and future research considerations are provided in Section \ref{sec:discussion}.

\section{Model description}\label{sec:model}

Let us denote by $\by\in\mathbb R^n$ the vector of responses $\by = (y_1,\ldots,y_n)^\top$ and $\X$ the  $n \times p$ matrix of covariates with elements $x_{ij}$, $i=1,\ldots,n$  and $j=1,\ldots,p$. The modeling framework under consideration assumes heterogeneity in the underlying distributions of the pairs $(y_i,\xxi^\top)$ across samples indexed by $i= 1,\ldots,n$ due to the presence of $K$ latent groups. We denote the group-specific parameters as $\thita = (\thitaY,\thitaX)^\top$ for $k=1,\ldots,K$, with $\thitaY$ and  $\thitaX$ corresponding to the associated parameters under group $k$ of the distributions of the response variable and the covariates, respectively. 
In addition, let $\z_i=(z_{i1},\ldots,z_{iK})^\top$ denote the vector of group indicators for $i=1,\ldots,n$ such that $z_{ik}\in\{0,1\}$ and $\sum_{k=1}^Kz_{ik}=1$.

If the number of groups $K$ and the true group allocations were known, then under group $k = 1,\ldots,K$ we have that
\begin{equation}
 p(y_i,\xxi | z_{ik}=1, \boldsymbol{\theta}_k)=
 p(y_i | \xxi, z_{ik}=1, \thitaY)
 p(\xxi| z_{ik}=1, \thitaX).
 \label{eq:known_k}
\end{equation}
In this work, we will assume Gaussian distributions for both parts in the right-hand side of \eqref{eq:known_k}. 
Specifically, the covariates are modeled via a $p$-dimensional multivariate normal distribution, meaning that  $\thitaX=(\mk,\mathrm{vec}(\Sk))^\top$, where $\mk$ is the vector of means, $\Sk$ the $p\times p$ covariance matrix, for cluster $k=1,\ldots,K$, and $\mathrm{vec}(\cdot)$ is the vectorization operator.
For the response variable, we assume a normal linear regression model within each cluster  $k=1,\ldots,K$ so that $\thitaY=(\ak,\bbk,\sk)^\top$, where $\ak$ represents the intercept term, $\bbk$ the vector of regression coefficients and $\sk$ the error variance. Given this formulation, the specific distributions in \eqref{eq:known_k} are as follows
\begin{align}
p(y_i| \xxi, z_{ik}=1, \thitaY) & \equiv     
f_{\mathrm{N}_1}(y_i|\ak+\xxti\bbk,\sk) \label{eq:normal_y}\\
 p(\xxi| z_{ik}=1, \thitaX) & \equiv  
f_{\mathrm{N}_p}(\xxi|\mk,\Sk),
\label{eq:normal_x}
\end{align}
where $f_{\mathrm{N}_q}(\cdot | \mathbf{m},\mathbf{\Sigma})$ denotes the density of the $q$-dimensional normal distribution with mean $\mathbf{m}$ and covariance matrix $\mathbf{\Sigma}$.
Below we present the resulting likelihood of the model under $K$ latent groups  as well as the complete prior specification of the proposed Bayesian CWM. 

\subsection{Likelihood}\label{sec:likelihood}

  Under the assumption of $K$ latent groups, each observation is \textit{a-priori} assigned to group $k$ with probability $\pi_k$, where $\sum_{k=1}^{K}\pi_k = 1$ and $0 < \pi_k\leqslant 1$, for $k=1,\ldots,K$. The vector $\bs{\pi}=(\pi_1,\ldots,\pi_K)^\top$ contains the weight of each group. The latent group indicator vectors $\z_i=(z_{i1},\ldots,z_{iK})^\top$ are, therefore, distributed as
\begin{align}
    \label{eq:z_prior}
    \z_i|\boldsymbol{\pi},K\sim\mathrm{Multinomial}(1|\pi_{1},\ldots,\pi_{K})
\end{align}
independently for $i=1,\ldots,n$, where  $\mathrm{Multinomial}(1|\pi_1,\ldots,\pi_K)$ denotes the multinomial distribution with one trial  and probability of category $k$ being equal to $\pi_k$, for  $k=1,\ldots,K$.
 
 Thus, given Equations \eqref{eq:known_k}, \eqref{eq:normal_y}, \eqref{eq:normal_x} and assuming i.i.d. samples within groups, the joint distribution of $(y_i, \xxi)$ conditional on $\boldsymbol{\theta}=(\boldsymbol{\theta}_1,\dots,\boldsymbol{\theta}_K)^\top$,  $\boldsymbol{\pi}$, $\z_i$ and $K$, is given by 
\begin{align*}
p(y_i,\xxi | \z_{i}, \boldsymbol{\theta},\boldsymbol{\pi}, K)&=\prod_{k=1}^{K}\left\{p(y_i|\thitaY,\xxi)p(\xxi|\thitaX)\right\}^{z_{ik}}\\ 
&= \prod_{k=1}^{K}\left\{f_{\mathrm{N}_1}(y_i|\ak+\xxti\bbk,\sk)f_{\mathrm{N}_p}(\xxi|\mk,\Sk)\right\}^{z_{ik}},\end{align*}
for  $i=1,\ldots,n$.
The marginal distribution of $(y_i,\xxi)$ is a mixture of $K$ distributions of the form
\begin{align*}
p(y_i,\xxi|  \boldsymbol{\theta},\boldsymbol{\pi}, K)  &=\sum_{k=1}^{K}\pi_k p(y_i|\thitaY,\xxi)p(\xxi|\thitaX)\nonumber\\ &= \sum_{k=1}^{K}\pi_k f_{\mathrm{N}_1}(y_i|\ak+\xxti\bbk,\sk)f_{\mathrm{N}_p}(\xxi|\mk,\Sk),\end{align*}
independently for $i=1,\ldots,n$. So, the observed likelihood is expressed as
\begin{align}
p(\by,\X|\boldsymbol{\theta},\boldsymbol{\pi}, K) & =\prod_{i=1}^{n}\sum_{k=1}^{K}
\pi_k p(y_i|\thitaY,\xxi)p(\xxi|\thitaX) \nonumber \\
& = \prod_{i=1}^{n}\sum_{k=1}^{K}
\pi_k f_{\mathrm{N}_1}(y_i|\ak+\xxti\bbk,\sk)f_{\mathrm{N}_p}(\xxi|\mk,\Sk)
\label{lik_marg}.
\end{align} 

We will work with the complete-data likelihood of the model in \eqref{lik_marg}, considering the latent group allocations as missing data. In this case, the {complete likelihood} can be expressed as
\begin{equation}
p(\by,\X,\mathbf{Z}|\boldsymbol{\theta},\boldsymbol{\pi})=
\prod_{i=1}^{n}\prod_{k=1}^{K}
\Big\{\pi_kf_{\mathrm{N}_1}(y_i|\ak+\xxti\bbk,\sk)f_{\mathrm{N}_p}(\xxi|\mk,\Sk)\Big\}^{z_{ik}},
\label{lik_full}
\end{equation}
where the dimensionality of $\mathbf{Z}$ is $n\times K$.   

\subsection{Prior distributions}\label{sec:prior}

In this section we describe the proposed prior specifications of the BGCW model. We begin with the prior distribution of $\boldsymbol{\pi}$ and the different approaches for handling an unknown number of $K$ components. Following this, we present the priors assigned to the parameters of the linear regression component and to the parameters of the multivariate predictor component of the proposed mixture model.  

\subsubsection{Prior over the mixing weights and approaches for unknown $K$}\label{sec:K}

For the conditional distribution of mixing proportions (given the number of components $K$), we make the usual assumption of a Dirichlet distribution, i.e., 
\begin{align}
\label{eq:mixing_proportions}
\boldsymbol{\pi}|\bs\gamma , K&\sim \mathrm{Dir}(\gamma_1,\ldots,\gamma_K),
\end{align}
where $\bs\gamma=(\gamma_1,\ldots,\gamma_K)$, with $\gamma_k > 0$ for all $k=1,\ldots,K$.
Usually, each component in a finite mixture model represents a cluster. However, this is not always the case. Another school of thought makes a clear distinction between the number of components in a mixture (that is, $K$) and the number of clusters, which corresponds to the number of non-empty components, i.e., components with at least one observation assigned to them. Below we review alternative approaches that we consider in our experimental results. 
In brief, when the number of components is considered fixed  we set $\gamma_k = 1$, for $k=1,\ldots,K$ (as is typical; see, e.g.,~\cite{marin2005bayesian} and Approach 1 below). In the  case of an unspecified number of components, we can estimate overfitting mixture models under a sparse Dirichlet prior (see Approach 2 below) or consider the more general approach where $K$ is random and fit a generalized mixture of finite mixture models (see Approach 3 below). 

\paragraph{Approach 1: Fixed number of components}
The simplest approach we could follow is to assume that the number of mixture components is fixed and each component represents a cluster. Then, we estimate models, separately, for a range of possible values $K\in\{1,\ldots,K_\mathrm{max}\}$, where $K_{\mathrm{max}}$ denotes a fixed constant. The number of components (or clusters) can then be selected according to  information criteria, such as the Akaike's Information Criterion \citep[AIC;][]{akaike_74}, the Bayesian Information Criterion \citep[BIC;][]{schwarz_78} or the Integrated Completed Likelihood  \citep[ICL;][]{biernacki2000assessing} criterion. Although this is not a fully Bayesian technique, we will consider it in our illustrations as a baseline approach due to its popularity within the model-based clustering community; see, e.g., the {\tt flexmix} \citep{flexmix0, flexmix1, flexmix2, flexmix3} and {\tt mclust} \citep{mclust} packages in  \textsf{R}. In this case, the Dirichlet concentration parameters $\gamma_1,\ldots,\gamma_K$ in Eq.~\eqref{eq:mixing_proportions} are all set equal to $1$, a choice leading to a uniform distribution defined in the set $\{\pi_k, k=1,\ldots,K: \pi_k > 0, \sum_{k=1}^{K}\pi_k=1\}$.

\paragraph{Approach 2: Overfitting mixture models} 

An overfitting mixture model is one where the number mixture components $K$ is larger that the ``true'' one, $K_+$, which represents the actual number of clusters in the data. \cite{rousseau2011asymptotic} showed that the posterior distribution of an overfitted mixture model, asymptotically, will empty the extra components after placing a sparse Dirichlet prior distribution in the mixing proportions.  Various implementations of this framework can be found in \cite{van2015overfitting, papastamoulis2018overfitting, papastamoulis2020clustering, papastamoulis2023model}. Under this setup, the number of mixture components is fixed to a constant value $K=K_{\mathrm{max}}$ which serves as an upper bound on the number of clusters. Then, inference for the number of clusters is based on the number of non-empty mixture components, i.e., components with at least one observation allocated to them, across the MCMC run. Regarding the Dirichlet concentration parameters $\gamma_1,\ldots,\gamma_K$ in Eq. \eqref{eq:mixing_proportions}, we follow Theorem 1 of \cite{rousseau2011asymptotic}, which requires that  $\max\{\gamma_k;k=1,\ldots,K_{\max}\} < d/2$ where $d$ denotes the number of free parameters of the distribution $p(y_i|\thitaY,\xxi)p(\xxi|\thitaX)$ in Equation \eqref{lik_marg}. Although this condition provides an upper bound, practical implementations typically adopt values of $\gamma_k$, $1,\ldots,K$, that are orders of magnitude smaller than $d/2$ to encourage emptying of  superfluous components (see, e.g., \cite{van2015overfitting, papastamoulis2018overfitting, papastamoulis2020clustering, papastamoulis2023model}). In the experiments presented below we set $K_\mathrm{max} = 20$ and $\gamma_k = 10^{-3}$, $k=1,\ldots,K_{\max}$. 


\paragraph{Approach 3: Generalized mixtures of finite mixtures}
Ultimately, the general framework of mixture models with a prior on the number of components is considered. We follow the approach in \cite{fruhwirth2021generalized}, modeling the data as a dynamic mixture of finite mixtures and estimating the posterior distribution using the telescoping sampler. The flexibility offered by this approach lies mainly on the fact that it explicitly samples the number of (unknown) mixture components, requiring only a relatively trivial MCMC step (unlike other approaches such as the reversible jump MCMC  sampler \citep{richardson1997bayesian}, for instance). 
In this method, a translated Beta-Negative Binomial $\mathrm{BNB}(a_\lambda,a_\pi,b_\pi)$ prior is placed on $K-1$, such that the corresponding probability mass function on the number of components is given by
\begin{equation}
    \label{eq:bnb_prior}
    p(K) = \frac{\Gamma(a_\lambda + K - 1)B(a_\lambda + a_\pi, K - 1 + b_\pi)}{\Gamma(a_\lambda)\Gamma(K)B(a_\pi, b_\pi)},
\end{equation}
where $\Gamma(\cdot)$ denotes the gamma function and $B(\cdot,\cdot)$ the beta distribution. Following \cite{fruhwirth2021generalized}, we set $a_\lambda = 1, a_\pi = 4, b_\pi = 3$, which results in a weakly informative prior with fat tails for the number of clusters. From Eqs. \eqref{lik_marg} and \eqref{eq:bnb_prior}, the joint marginal distribution of the data $(\by, \X)$ is represented as a countably infinite mixture of finite mixtures with $K$ components
\[
p(\by, \X|\{\bs\pi_K, \bs\theta_K\}_{K=1}^{\infty}) = \sum_{K=1}^{\infty}p(K)\prod_{i=1}^{n}\sum_{k=1}^{K}\pi_{k,K} p(y_i|\bs\theta_{k,K}^{\scaleto{Y}{4pt}},\bs x_i)p(\bs x_i|\bs\theta_{k,K}^{\boldsymbol{{\scaleto{X}{4pt}}}}),
\]
where $\bs\pi_K := (\pi_{1,K},\ldots,\pi_{K,K})$ and $\bs\theta_K := (\{\bs\theta_{1,K}^{\scaleto{Y}{4pt}}\bs\theta_{1,K}^{\boldsymbol{{\scaleto{X}{4pt}}}}\},\ldots,\{\bs\theta_{K,K}^{\scaleto{Y}{4pt}}\bs\theta_{K,K}^{\boldsymbol{{\scaleto{X}{4pt}}}}\})$, for $K\in \mathbb{N}$.
Finally, following the dynamic mixture of finite mixtures approach \citep{fruhwirth2021generalized}, we set the concentration parameters in Eq.~\eqref{eq:mixing_proportions} as 
\begin{equation}
    \label{eq:common_gamma}
    \gamma_1 = \cdots=\gamma_K = \frac{\gamma}{K},
\end{equation}
placing Snedecor's F distribution as a hyper-prior distribution on $\gamma$ 
\begin{equation}
    \label{eq:dynamic_a_prior}
    \gamma\sim\mathrm{F}(\nu_l,\nu_r),
\end{equation}
and setting $\nu_\ell = 6$ and $\nu_r = 3$. 


\subsubsection{Priors for the linear regression mixture component}\label{sec:reg}      

The linear regression component-specific parameters are assumed to be independent conditional on $K$. Specifically, we consider the following prior design for subgroups indexed by $k=1,\ldots,K$:
\begin{align}
\ak|K & \sim \mathrm{N}(0, \sigma_{\alpha}^2), \label{eq:intercept}\\
\beta_{kj}|\sk,\lk, K & \sim \mathrm{DE}\Bigg(0,\displaystyle\frac{\lambda_k}{\sigma_k}\Bigg)\label{double-exp},\\
\sigma_k^2|K  & \sim \mathrm{IG}(a,b)\label{inv-gamma}, \\
\lambda_k|K & \sim \text{half-Cauchy}(0,1),\label{half-cauchy-dist}
\end{align}
for $j=1,\ldots,p$, where $\mathrm{DE}(\chi,\rho)$ in \eqref{double-exp} denotes the double exponential distribution with location $\chi$ and {rate} $\rho$, and $\mathrm{IG}(\chi,\rho)$ in \eqref{inv-gamma} the inverse-gamma distribution with shape $\chi$ and rate $\rho$. 


The normal  prior for the intercepts in  \eqref{eq:intercept}, centered at zero with large variance (we set $\sigma_{\alpha}^2=10^3$), is a standard vague prior for regression analysis. The double-exponential shrinkage prior for the regression coefficients in \eqref{double-exp} is according to the Bayesian lasso formulation of \cite{park_casella}; specifically, we have that
\begin{equation}
p(\bbk|\sk,\lk, K)=\prod_{j=1}^p p(\beta_{kj}|\sk,\lk) = \displaystyle \prod_{j=1}^p 
\frac{\lk}{2\sigma_k}
\exp\Bigg(-\lk\displaystyle\frac{\vert\beta_{kj}\vert}{\sigma_k}\Bigg).
\label{lasso-prior}
\end{equation}
As discussed in \cite{park_casella} the prior in Eq. \eqref{lasso-prior} can be written as a scale mixture of normal distributions with an exponential mixing density, leading to an augmented ``global-local'' shrinkage-prior representation of the following form
\begin{align}
p(\bbk|\sk,\tau^2_{k1},\ldots,\tau^2_{kp}, K) & \sim\mathrm{N}_p\big(\mathbf{0}_p,\sk\Tk\big), \mbox{~with~}\Tk =\mathrm{diag}(\tau^2_{k1},\ldots,\tau^2_{kp}), \label{aug_beta}\\
\tau^2_{kj}|K &\sim \mathrm{Exp}(\lambda_k^2/2), ~ \mbox{for} ~ j=1,\ldots,p. \label{aug_tau}
\end{align} 
In Equation \eqref{aug_tau},  the general notation $\mathrm{Exp}(\lambda)$  refers to the exponential distribution with {rate} $\lambda > 0$. The above augmentation facilitates the use of Gibbs sampling. 
The inverse-gamma prior in \eqref{inv-gamma}, whose limiting form is Jeffreys prior for $a\rightarrow 0$ and $b\rightarrow 0$ (we use  $a=b=10^{-3}$), is the standard option for variance components, leading to conditional conjugacy.

The prior formulation for the regression part of the model is completed by the assignment of a half-Cauchy prior on the group-specific penalty parameters $\lk$; specifically, from \eqref{half-cauchy-dist}, we have that
\begin{equation}
p(\lambda_k|K)=\displaystyle\frac{2}{\pi(1+\lambda_k^2)}.
\label{hCauchy}
\end{equation}
 The use of half-Cauchy priors for scale parameters of this type is justified in the relevant work of \cite{Polson_Scott_2012};  namely, based on the argument that the prior density does not vanish at the origin (as it would be the case with an inverse-gamma prior), constituting half-Cauchy priors a common choice for penalty parameters \citep[see, e.g.,][]{review_shrinkage}. In addition, one can make use of a convenient augmentation scheme in order to facilitate  Gibbs sampling, expressing the half-Cauchy distribution in \eqref{hCauchy} as a scale mixture of a half-normal with a gamma mixing density; namely,
 \begin{align}
\lk|\delta_k, K & \sim \text{half-Normal}(0,\delta_k^{-1}), \label{aug_lambda}\\
\delta_k|K & \sim \mathrm{Gamma}(1/2,1/2), \label{aug_delta}
 \end{align}
 so that $p(\lk|\delta_k, K)~\propto~ \delta_k^{1/2}\exp\big(-\delta_k\lambda_k^2/2\big)$ and $p(\delta_k|K)~\propto~ \delta_k^{1/2-1}\exp(-\delta_k/2)$,
 with the integration $\int p(\lk|\delta_k, K)p(\delta_k|K)\mathrm{d}\delta_k$ resulting in the distribution in \eqref{hCauchy}. Note that the symbol $\propto$ means ``proportional to''.
The resulting full conditional distributions, used for Gibbs sampling, are presented in Section \ref{sec:gibbs}. 

\subsubsection{Priors for the Gaussian covariate mixture component}\label{sec:x}

For the second component of the model that takes into account the distribution of covariates we begin by specifying multivariate normal priors for the group-specific means conditional on the group-specific covariances, so that 
\begin{equation}
\mk | \Sk, K \sim \mathrm{N}_p(\mathbf{m}_0, \Sk), 
\end{equation}
independently for $k =1,\ldots,K$. Above, without loss of generality, we can assume that $\mathbf{m}_0=\mathbf{0}_p$ in absence of prior knowledge. For the group-specific precision matrices $\Sk^{-1}=\Ok$, we adopt the Bayesian graphical lasso approach \citep{Wang2012} which entails using independent double-exponential distributions for the off-diagonal terms and independent exponential distributions for the elements of the main diagonal; specifically, we have that 
\begin{equation}
\Ok =\Sk^{-1}: p(\Ok|\psi_k,K) \propto 
\bigg[
\prod_{j=1}^pf_\mathrm{Exp}(\omega_{kjj}|\psi_k/2)
\prod_{j<l,l=2}^p
f_\mathrm{DE}(\omega_{kjl}|0,\psi_k)
\bigg]\mathbbm{1}_{(\Ok\in M^+)},
\label{glasso}
\end{equation}
where $\psi_k > 0$ is the corresponding graphical lasso penalty, $\mathbbm{1}_{(\cdot)}$ is the indicator function  and $M^+$ is the space of positive definite matrices. Above, $f_\mathrm{Exp}(\cdot| q)$ and $f_\mathrm{DE}(\cdot| 0,q)$ denote the respective densities of the exponential distribution with {rate} $q$ and of the zero-mean double exponential distribution with {rate} $q$. 
Here again, in order to facilitate posterior simulation via Gibbs sampling, instead of Eq. \eqref{glasso}, one can use the following expansion   
\begin{align}
 p(\Ok|\psi_k,\boldsymbol{\phi}_k,K) & ~ \propto ~  
\bigg[
\prod_{j=1}^pf_\mathrm{Exp}(\omega_{kjj}|\psi_k/2)
\prod_{j<l,l=2}^p
f_{\mathrm{N}_1}(\omega_{kjl}|0,\phi_{kjl})
\bigg]\mathbbm{1}_{(\Ok\in M^+)}, \label{aug_glasso}\\
\phi_{kjl}|K & \sim \mathrm{Exp}(\psi_k^2/2), 
\label{aug_phi}
\end{align}
for $j<l$ and  $l=2,\ldots,p$. 
Finally, the prior formulation for this part of the model is completed via the specification of a gamma hyper-prior for $\psi_k$; namely, 
\begin{equation}\label{eq:psi}
\psi_k|K \sim \mathrm{Gamma}(r,s),
\end{equation}
independently for $k=1,\ldots,K$, with the default values $r=1$ and $s=0.01$ used in \cite{Wang2012}. The corresponding full conditional distributions, from the above prior specification, are presented below, while the complete MCMC algorithm can be found in Appendix A.


\section{Posterior inference}\label{sec:MCMC}

\subsection{MCMC sampling for fixed number of components}\label{sec:gibbs}

Naturally, the resulting posterior distribution is intractable. However, under the prior specifications in Section \ref{sec:prior} and, especially, by making use of the augmentation schemes therein, it is possible to fully derive all the full conditional distributions required for Gibbs sampling.

Specifically, the full conditional distribution of the latent indicators $z_{ik}$ is a multinomial distribution with success probabilities $p_{ik}~\propto~ f_{\mathrm{N}_1}(y_i|\ak+\xxti\bbk,\sk)f_{\mathrm{N}_p}(\xxi|\mk,\Sk)\pi_k$ for $i=1,\ldots,n$ and $k=1,\ldots,K$. For the linear regression component of the mixture we have 
\begin{align*}
  \ak | \cdot & \sim\mathrm{N}_1\big(n_k^{-1}w_k(\by-\X\bbk)^\top\z_k,n_k^{-1}w_k\sigma_k^2\big)\\
  \bbk| \cdot & \sim\mathrm{N}_p\big(\Ak^{-1}\X^\top\mathbf{Z}_k(\by-\ak\mathbf{1}_{n},\sigma_k^2\Ak^{-1})\big)\\
  \sigma_k^2 |\cdot & \sim \mathrm{IG}
\left(a + \frac{1}{2}(n_k+p),
b+\frac{1}{2}[\mathbf{e}_k^\top\mathbf{Z}_k\mathbf{e}_k + \bbk^\top\Tk^{-1}\bbk]\right)
\end{align*}
where $n_k=\sum_i  \mathbbm{1}_{(z_{ik}=1)}$, $\z_k=(z_{1k},\ldots,z_{nk})^\top$, $\mathbf{Z}_k=\mathrm{diag}(\z_k)$, $w_k= \sigma_{\alpha}^2/(\sigma_{\alpha}^2 +n_k^{-1}\sigma_k^2)$, $\Ak = (\X^\top\mathbf{Z}_k\X + \Tk^{-1})$ and 
$\mathbf{e}_k=\by-\ak\mathbf{1}_{n}-\X\bbk$. The general notation $x|\cdot$ is used to indicate the conditional distribution of a random variable $x$ amongst a set of random variables. Furthermore, from the augmentation schemes  presented in Equations \eqref{aug_beta}, \eqref{aug_tau}, \eqref{aug_lambda} and \eqref{aug_delta} we obtain 
\begin{align*}
 1/\tau^2_{kj}|\cdot & \sim \mathrm{InvGaussian}\big(\sigma_k\lambda_k\beta_{kj}^{-1},\lambda_k^2\big)\\
 \lambda_k^2 |\cdot & \sim \mathrm{Gamma}\left(p+0.5, 0.5\left(\sum_{j=1}^p \tau_{kj}^2+\delta_k\right)\right) \\
 \delta_k|\cdot & \sim \mathrm{Gamma}\big(1, 0.5(\lambda_k^2+1)\big).
\end{align*}

Continuing, for the multivariate Gaussian component of the mixture model we initially get that $\mk | \cdot \sim \mathrm{N}_p \big((n_k+1)^{-1}(\mathbf{m}_0+\sum_iz_{ik}\xii),(n_k+1)^{-1}\Sk\big)$. Sampling $\Ok$ and $\boldsymbol{\phi}_k$ from their full conditionals directly, based on the priors in \eqref{aug_glasso} and \eqref{aug_phi}, is not possible; however, as shown in \cite{Wang2012} it is possible to use a block Gibbs sampling scheme, based on reparametrizations, which produces row/column-wise updates of $\Ok$. The scheme is described and discussed in detail in \cite{Wang2012}; here we will provide a brief description. First for each row and column $j=1,\ldots,p$ the following partitions are defined
\begin{equation}
\boldsymbol{\Omega}_{k}^{(j)} = \begin{bmatrix}
\boldsymbol{\Omega}_k^{(\smallsetminus j,\smallsetminus j)} & \boldsymbol{\omega}_k^{(\smallsetminus j,j)} \\
\boldsymbol{\omega}_k^{(\smallsetminus j,j)^\top} &
\omega_{kjj}
\end{bmatrix}, ~
\mathbf{S}_k^{(j)} = \begin{bmatrix}
\mathbf{S}_k^{(\smallsetminus j,\smallsetminus j)} & 
\mathbf{s}_k^{(\smallsetminus j,j)} \\
\mathbf{s}_k^{(\smallsetminus j,j)^\top} &
s_{kjj}
\end{bmatrix} ~
, \mathbf{\Phi}_{k}^{(j)}  = \begin{bmatrix}
\boldsymbol{\Phi}_k^{(\smallsetminus j,\smallsetminus j)} & \boldsymbol{\phi}_k^{(\smallsetminus j,j)} \\
\boldsymbol{\phi}_k^{(\smallsetminus j,j)^\top} &
0
\end{bmatrix},
\label{partitions}
\end{equation}
where for a $n\times n$ symmetric matrix $\mathbf{Q}$, $\mathbf{Q}^{(\smallsetminus i,\smallsetminus i)}$ is the $(n-1) \times (n-1)$ matrix without the $i$-th row/column of $\mathbf{Q}$ and $\mathbf{q}^{(\smallsetminus i, i)}$ is the $i$-th column of $\mathbf{Q}$ without element $q_{ii}$. In addition, we have $\mathbf{S}_k=\sum_i z_{ik}(\xxi-\mk)(\xxi-\mk)^\top+(\mk-\mathbf{m}_0)(\mk-\mathbf{m}_0)^\top$, while $\mathbf{\Phi}_{k}$ is a symmetric matrix with zeros in the main diagonal and the elements of $\boldsymbol{\phi}_k$ filling the upper diagonal entries. Having defined the above partitions we sample  $\eta_{k1} | \cdot\sim \mathrm{Gamma}\big(0.5(n_k+1), 0.5(s_{kjj}+\psi_k)\big)$,  $\boldsymbol{\eta}_{k2}|\cdot\sim\mathrm{N}_{p-1}\big(-\mathbf{C}\mathbf{s}_k^{(\smallsetminus j,j)},\mathbf{C}\big)$, where $\mathbf{C}= \big((s_{kjj}+\psi_k)\mathbf{\Omega}_{k}^{(\smallsetminus j,\smallsetminus j)^{-1}}+\mathrm{diag}(\boldsymbol{\phi}_k^{(\smallsetminus j,j)})^{-1}\big)$, and then set $\boldsymbol{\omega}_k^{(\smallsetminus j,j)} =\boldsymbol{\eta}_{k2}$, and $\omega_{kjj}= \eta_{k1}+ \boldsymbol{\eta}_{k2}^\top\mathbf{\Omega}_{k}^{(\smallsetminus j,\smallsetminus j)^{-1}} \boldsymbol{\eta}_{k2}$, for $j=1,\ldots,p$. Subsequently, for $j<l ~ (l = 2,\ldots,p),$ we draw $u_{kjl}\sim\mathrm{InvGaussian}\big(\omega_{kjl}^{-1}\psi_k,\psi_k^2\big)$, and set $\phi_{kjl}=u_{kjl}^{-1}$.
The graphical lasso penalty is updated from its corresponding full conditional, which is $\psi_k| \cdot\sim \mathrm{Gamma}\big(r+0.5p(p+1),s+0.5\Vert\Ok\Vert_1\big)$, where $\Vert\cdot\Vert_1$ denotes the $\ell_1$ norm. 

Finally, the full conditional distribution of mixing proportions is simply $\boldsymbol{\pi}|(\bs z, K,\gamma)\sim \mathrm{Dir}(\gamma_1+n_1,\ldots,\gamma_K+n_K)$. 



\subsection{Sampling an unknown number of components}\label{sec:telescoping}

In the case of an unknown number of components, the MCMC sampler under Approach 3 in Section \ref{sec:prior} should also update $K$.  Under the prior in Equation \eqref{eq:bnb_prior}, the full conditional distribution of the number of components is given by \citep{fruhwirth2021generalized}
\begin{equation}
    \label{eq:k_conditional}
    p(K|\bs z, \gamma_K)~\propto ~p(K)\frac{K!}{(K-K_+)!}\frac{\Gamma(\gamma_K K)}{\Gamma(n+\gamma_KK)\Gamma(\gamma_K)^{K_+}}\prod_{j:n_j > 0}\Gamma(n_j + \gamma_K),
\end{equation}
for $K\in \{K_+,K_++1,\ldots\}$, where $K_+$ denotes the number of non-empty components (clusters) induced by $\bs z$.
The hyperparameter $\gamma$ is updated using a Metropolis-Hastings step in order to sample from the corresponding conditional distribution, that is,
\begin{equation}
\label{eq:alpha_conditional}
p(\gamma|\bs z, K, \nu_\ell, \nu_r) ~ \propto ~  p(\gamma)\frac{\gamma^{K_+}\Gamma(\gamma)}{\Gamma(n+\gamma)}\prod_{k=1}^{K_+}\frac{n_k+\gamma/K}{\Gamma(1+\gamma/K)},
\end{equation}
as described in \cite{fruhwirth2021generalized}. 
The main steps of the proposed telescoping sampler are summarized in the form of a pseudocode in Section A of the Appendix.



%

\subsection{Identifiability issues}\label{sec:identifiability}

Mixtures of distributions impose various identifiability concerns, including trivial and generic identifiability \citep{fruhwirth2006finite}. The term trivial identifiability refers to problems related to the invariance of the likelihood with respect to permutations of the parameters, i.e., the  label-switching problem \citep{redner1984mixture}. This can also include cases with empty components and the presence of components with identical parameters \citep{grun2008identifiability}. We deal with label switching applying the equivalence classes representatives (ECR) algorithm \citep{papastamoulis2010artificial, papastamoulis2016} in order to post-process the generated MCMC samples and derive meaningful estimates of the marginal posterior distributions. The ECR algorithm uses the generated allocation variables of the mixture model and by default acts to the occupied mixture components in the MCMC output, conditional on the number of components. 

Generic identifiability refers to the question of what can be estimated consistently. This issue has been addressed in \cite{hennig2000identifiablity} for mixtures of Gaussian linear regressions with fixed or random covariates. In the case of random covariates, the main result requires a check of a coverage condition in order to ensure generic identifiability; specifically, that the distribution of the covariates does not assign positive probability to any $(p-1)$-dimensional hyper-plane \citep[see Theorem 3.2 in][]{hennig2000identifiablity}, where $p$ denotes the number of covariates. In general, caution is needed when checking the validity of this condition in cases where discrete covariates are considered. However, in our setup, this condition is satisfied as the conditional distribution of covariates within each cluster is a $p$-dimensional multivariate normal distributions.

\subsection{Post-hoc variable evaluation}
\label{sec:var_sel}

Following \cite{Papastamoulis2022}, we assess variable significance, ex-post by estimating simultaneous credible regions of the regression coefficients. We apply this procedure cluster-wise, after post-processing the retained MCMC sample in order to undo label-switching. In case where a zero is not contained in the simultaneous credible region for at least one cluster, the corresponding variable is deemed as ``significant''. On the other hand, if a zero is contained in the simultaneous credible region for all clusters, the corresponding variable is deemed as ``non-significant''.  This scheme is not a strictly Bayesian variable selection approach; however, it is straightforward to apply within our model and leads to reliable results according to the simulation studies presented next.

We denote by $\boldsymbol{I}_k = I_{k1}\times \cdots\times I_{kp}$ the (estimated) $100(1-\alpha)\%$ simultaneous credible region of $\bs\beta_{k}=(\beta_{k1},\ldots,\beta_{kp})^\top$, for cluster $k = 1,\ldots,K$. We then define the binary vector $\hat{\bs\xi}_k=(\hat\xi_{k1},\ldots,\hat\xi_{kp})^\top$, where 
\[
\hat\xi_{kj} = \begin{cases}
    1, &\mbox{if}\quad 0\notin I_{kj}\\
    0, &\mbox{otherwise.}
\end{cases}.
\]
 The subset $S_k=\{j=1,\ldots,p: \hat\xi_{kj} = 1\}$ contains the variables deemed as significant for cluster $k$. Finally, we define the binary vector $\hat{\bs\xi} = (\hat\xi_1,\ldots,\hat\xi_p)^\top$ where 
 \begin{equation}
 \label{eq:vs}
  \hat\xi_j = 1-\prod_{k=1}^{K}(1 - \hat{\xi}_{kj})   
 \end{equation}
 for $j = 1,\ldots,p$. The subset $S=\{j=1,\ldots,p: \hat\xi_j = 1\}$ contains the variables deemed significant in at least one cluster.

\section{Simulation studies}\label{sec:sim}

We evaluate the performance of the proposed method in synthetic datasets. In Section \ref{sec:cwm_sim} the synthetic data is generated from the cluster weighted model, under four distinct scenarios corresponding to different variability patterns of the covariates. In order to evaluate the robustness of the proposed approach in cases where the cluster weighted model is not the ``true'' one, we also consider a scenario where data are generated from a mixture of experts in  Section \ref{sec:moe_sim}. 

Comparisons are made against {\tt flexCWM} \citep[version 1.92,][]{flexcwm},  {\tt FLEXMIX} \citep[version 2.3-19,][]{flexmix0}, 
{\tt MoEclust}  \citep[version 1.5.2,][]{moeclust} 
and {\tt RJM} \citep[Regularized Joint Mixtures, see][]{perrakis2023regularized} implementations. As noted in Section \ref{intro}, these methods primarily differ in their treatment of covariates. \texttt{FLEXMIX} and \texttt{MoEClust} treat covariates as fixed, whereas cluster-weighted models (\texttt{flexCWM}, \texttt{RJM}) and the proposed \texttt{BGCWM} treat them as random. For cluster-weighted models, \texttt{flexCWM} achieves sparsity by using parsimonious covariance matrix representations, whereas \texttt{RJM} and \texttt{BGCWM} enforce it indirectly through shrinkage prior distributions.
 In addition, \texttt{FLEXMIX} and \texttt{MoEClust} can incorporate concomitant variables, allowing covariates to influence the mixture proportions. The results presented below for \texttt{FLEXMIX} and \texttt{MoEClust} correspond to those from the selected model  (either with or without concomitant variables) based on the respective information criterion. The reader is referred to Section F of the Appendix for more detailed descriptions of these competing approaches.

\subsection{Simulations from a cluster weighted model}\label{sec:cwm_sim}

Synthetic datasets were generated through simulations with the true number of clusters set to \( K \in \{2, 3, 4\} \). The number of predictors and sample sizes varied across \( p \in \{9, 18\} \) and \( n \in \{500, 1000\} \), respectively. 
In order to incorporate various unbalanced cluster sizes, the mixing proportions are defined as $\pi_k~\propto~ k$, for $k=1,\ldots,K$. 

The parameters related to the linear regression mixture components were generated as
\begin{align*}
\alpha_{k}&\sim\mathrm N(0,10^2)\\
\beta_{kj}&\sim p_0 1_{\{0\}}  + (1-p_0)\mathrm N(0, 3^2)\\
\sigma^2_k&\sim\mathrm N(0,1)
\end{align*}
independently for  $k = 1,\ldots,K$ and $j = 1,\ldots,p$, where $1_{\{0\}}$ denotes a degenerate distribution at $0$. We further consider two cases of sparsity, using $p_0 = 2/3$ and $p_0 = 1/3$. Table \ref{tab:influential-vars} summarizes the expected number of influential (non-zero) coefficients per cluster.

 \begin{table}[h]
    \centering
    \begin{tabular}{ccccc}
        \toprule
        Sparsity level ($p_0$) & & $2/3$ & $1/3$ \\
        \midrule
        \multirow{2}{*}{Number of variables ($p$)}
        & $9$  & 3  & 6  \\
        & $18$ & 6  & 12 \\
        \bottomrule
    \end{tabular}
    \caption{Expected number of influential variables per cluster.}
    \label{tab:influential-vars}
\end{table}

For the simulation of the covariate matrices we consider the following four scenarios across clusters. 
\begin{enumerate}
\item \textbf{Uncorrelated homogeneous covariates.}
The covariates within each cluster are independent with unit variance, i.e., $\bs\Sigma_k = \mathbf{{I}}_p$, for $k = 1,\ldots,K$ and $\bs\mu_k = \bs 0_p$, where $\bs 0_p = (0,\ldots,0)^\top$. In this case, the covariates are indistinguishable among clusters, since  $\bf X$ and the latent allocation matrix $\bf Z$ are independent.

\item \textbf{Correlated homogeneous covariates.}
The covariates within each cluster have common variance $\bs\Sigma_k = \bs\Sigma$, for $k = 1,\ldots,K$, where $\bs\Sigma$ is generated from a Wishart distribution, i.e., $\bs\Sigma \sim \mathcal W_p(\bf V, 20)$. A Toeplitz structure is used for the positive-definite scale matrix $\bf V$, so that  the entries of the matrix are constant along each diagonal and symmetric. In particular, we used $ V_{ij} = (p-|i-j|)/p$, for $i,j=1,\ldots,p$. Setting $\bs\mu_k = \bs 0_p$ results again in indistinguishable covariates across clusters.

\item \textbf{Uncorrelated heterogeneous covariates.}
The covariates within each cluster are 
$\bs\Sigma_k = {\bf {I}}_p$, for  $k = 1,\ldots,K$. 
The mean for cluster $k$ is set equal to 
\[\mu_{kj} = \tilde\mu_{\rho_k j},\]
where $\rho = (\rho_1,\ldots,\rho_K)$ is a random permutation of $\{1,\ldots,K\}$ and for $j=1,\ldots,p$:
\[\tilde\mu_{kj} =  \begin{cases}
    0, &\quad k =1 \\
    2\sin\left\{(j-1)k\pi/p\right\}, &\quad k = 2\\
    2\cos\left\{(j-1)k\pi/p\right\}, &\quad k = 3\quad (\mbox{if $K\in\{3,4\}$})\\
    4\sin^2\left\{(j-1)k\pi/p\right\} - 4\cos^2\left\{(j-1)k\pi/p\right\}, &\quad k = 4\quad (\mbox{if $K = 4$})\\
\end{cases}\]

\item \textbf{Correlated heterogeneous covariates.}
The covariates are generated from a mixture of Gaussian factor analyzers \citep[see][]{papastamoulis2018overfitting, papastamoulis2020clustering}, where each cluster exhibits distinct covariance structure $\bs \Sigma_k$ and $\bs\mu_k$ as in Scenario 3.
\end{enumerate}
For each combination of number of clusters ($K$), number of variables ($p$), sample size ($n$), covariate scenario and sparsity levels ($p_0$), 30 replicated datasets were generated ($3\times 2\times 2\times 4\times 2\times 30 = 2880$ datasets in total) based on Eq. \eqref{lik_marg}.  Technical details related to initialization of the algorithm, MCMC and post-processing are presented in Appendix B.
We consider three inferential tasks: (i) estimation of the number of clusters, (ii) clustering performance and (iii) post-hoc variable evaluation. 
    \label{fig:overall}

\begin{figure}[p]
    \centering
    \includegraphics[scale = 0.55]{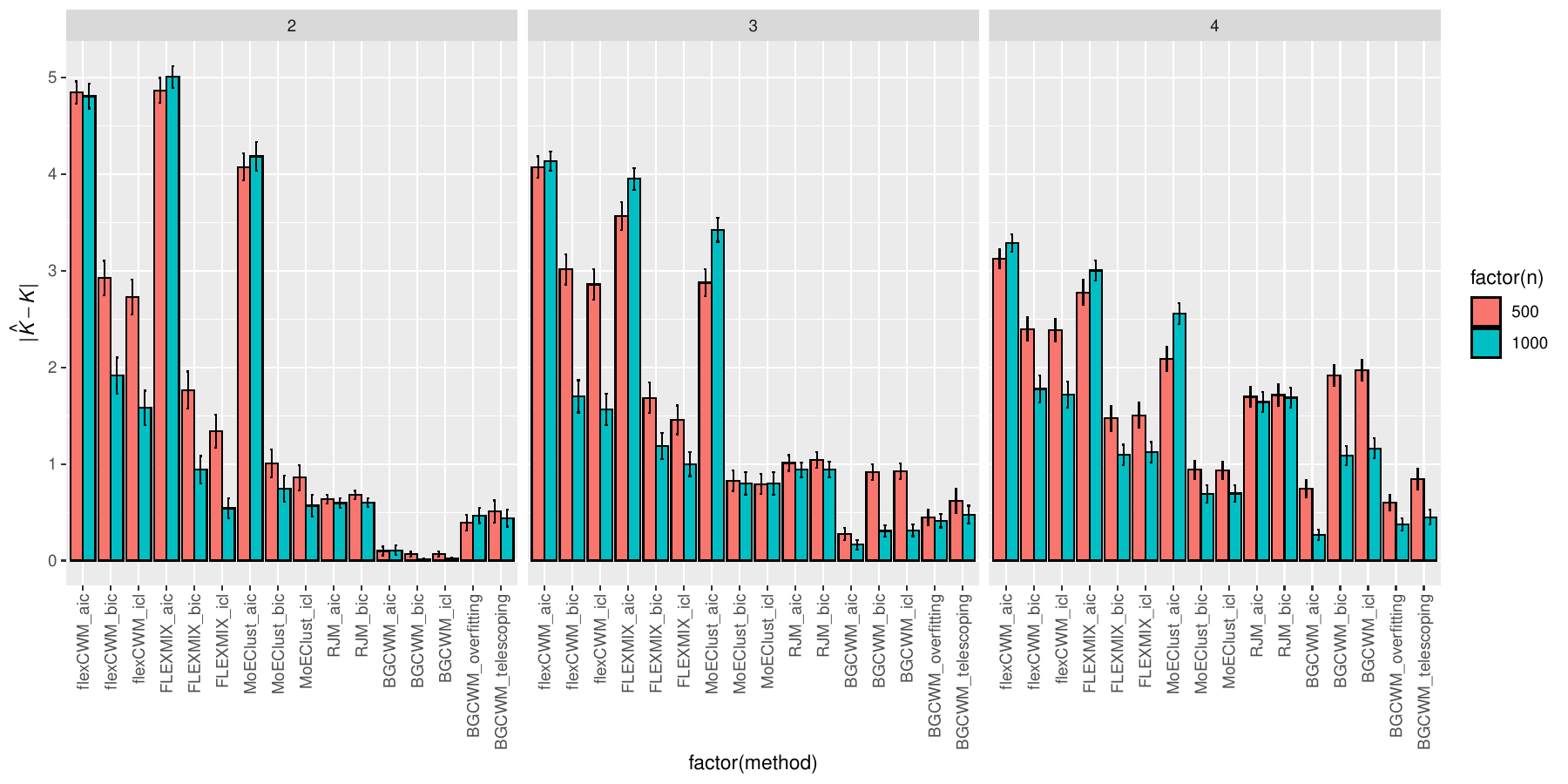}
    \caption{Mean absolute error of estimating the number of clusters for all simulation scenarios per value of $K\in\{2, 3, 4\}$. The error bars display (asymptotic) $95\%$ confidence intervals of the mean.}
    \label{fig:overall1}
\end{figure}

Overall results (across all covariate-simulation scenarios) related to the estimation of $K$,  are summarized in Figure \ref{fig:overall1} 
in terms of the mean absolute error $|K - \hat K|$, where $K$ and $\hat K$ denote the true and estimated number of clusters, respectively. 
MCMC estimation based on information criteria ({\tt BGCWM\_aic}, {\tt BGCWM\_bic}, {\tt BGCWM\_icl}) performs well for $K=2$. However, as $K$ increases (especially when $K=4$) the performance under {\tt BGCWM\_bic} and {\tt BGCWM\_icl} deteriorates; this is generally expected as these criteria impose stronger penalties on large models. The overfitting mixture ({\tt BGCWM\_overfitting}) and the telescoping sampler ({\tt BGCWM\_telescoping}) produce similar results. When $K = 2$ the information criteria seem to outperform these methods, however, for larger values of $K$ (particularly for $K=4$) the overfitting mixture model and the telescoping sampler yield similar or better performance in comparison to information criteria. EM-based RJM approaches ({\tt RJM\_aic} and {\tt RJM\_bic}) produce inferior results in all cases and the same holds for {\tt flexCWM}, {\tt FLEXMIX} and {\tt MoEclust}. A breakdown of the results per covariate simulation scenario is presented in Appendix C (Figure C.1).

\begin{figure}[p]
    \centering
     \includegraphics[scale = 0.55]{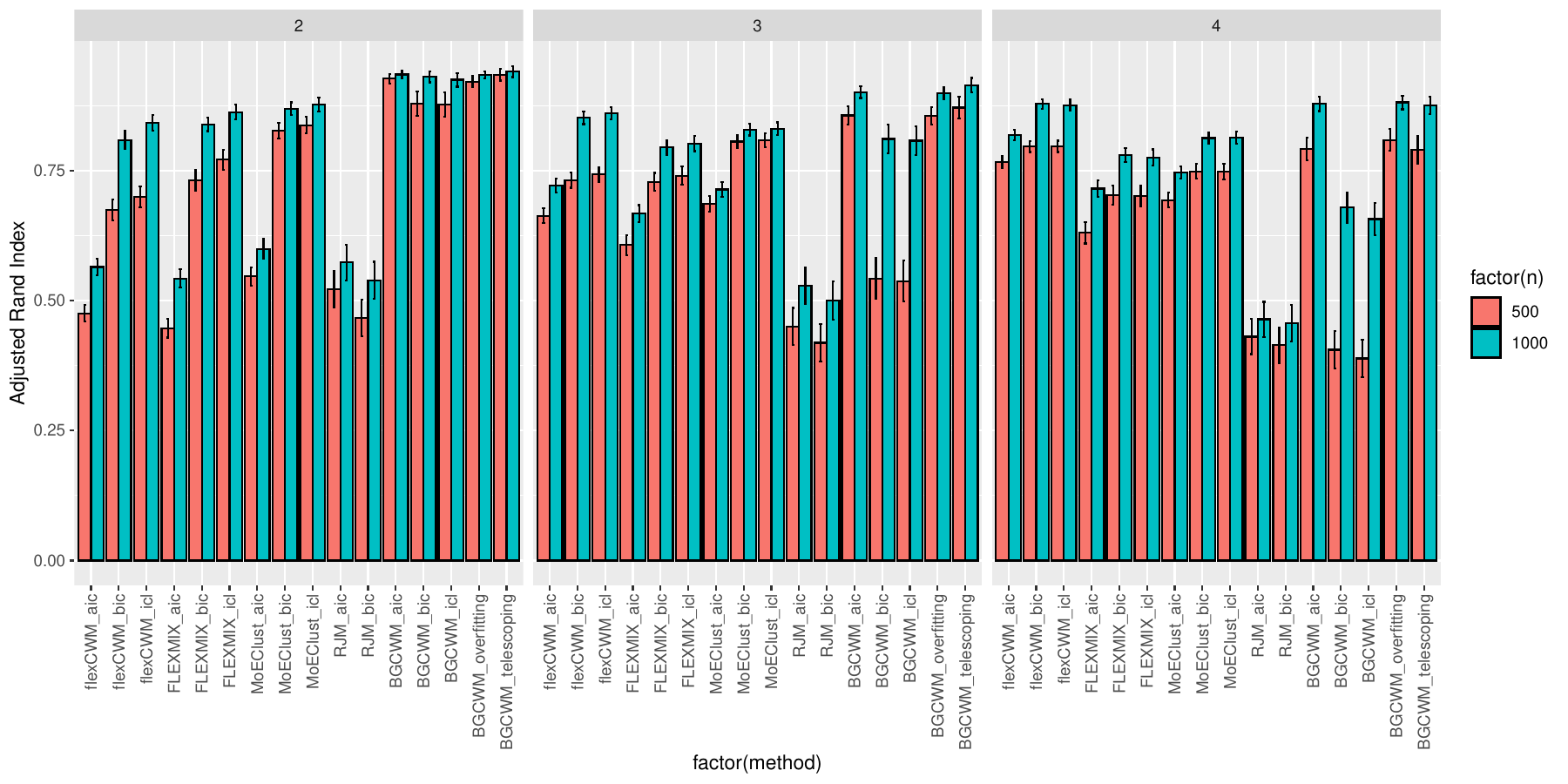}
     \caption{Mean adjusted Rand index for all simulation scenarios per value of $K\in\{2, 3, 4\}$. The error bars display (asymptotic) $95\%$ confidence intervals of the mean.}
    \label{fig:overall2}
\end{figure}

Clustering performance is quantified in terms of the adjusted Rand index. In Figure \ref{fig:overall2} 
we observe that the best performance is obtained from {\tt BGCWM\_aic}, {\tt BGCWM\_overfitting} and {\tt BGCWM\_telescoping}, followed by {\tt flexCWM}, {\tt MoEclust\_bic}, {\tt MoEclust\_icl}, {\tt FLEXMIX\_bic} and {\tt FLEXMIX\_icl} methods. Results per covariate simulation scenario can be found in Appendix C (Figure C.2).

\begin{figure}[h]
    \centering
    \includegraphics[scale = 0.52]{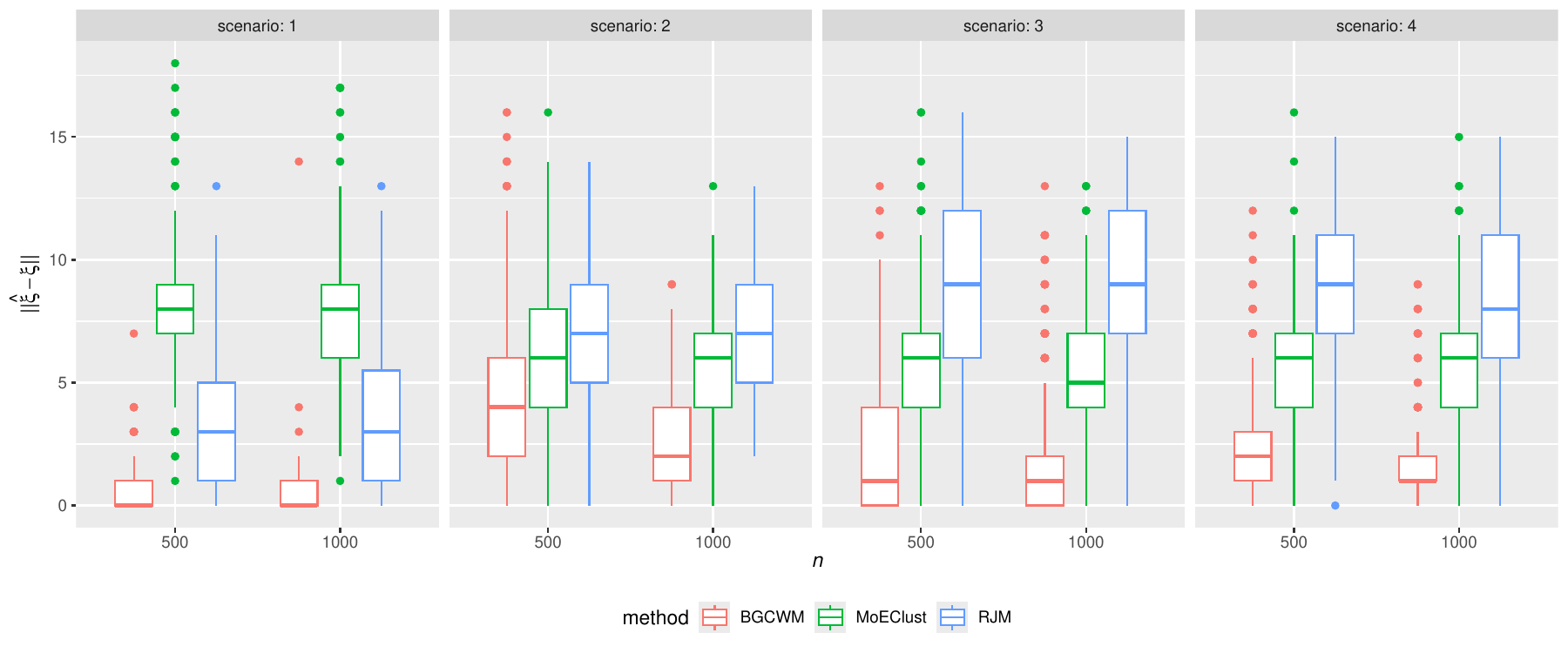}
    \caption{Boxplots of the number of falsely selected/not selected variables per covariate-simulation scenario and sample size $n\in\{500, 1000\}$  (the number of covariates $p$  is equal to 18).}
    \label{fig:variable_selection}
\end{figure}

Lastly, we examine variable significance based on critical regions.  
We assess selection accuracy by reporting the number of falsely selected or falsely not selected variables with respect to the ground truth, that is, 
$
||\hat{\bs\xi} - \bs\xi|| = \sum_{j=1}^{p}|\hat\xi_{j} - \xi_j|
$,
where the vector $\hat{\bs\xi}$ is defined in Eq. \eqref{eq:vs}. The results are summarized in Figure \ref{fig:variable_selection} where we compare {\tt BGCWM\_telescoping} against {\tt MoEClust} and the lasso implementation of {\tt RJM}. Each panel shows the number of falsely selected or not selected variables per simulation scenario (1, 2, 3, 4) for different sample sizes $n\in\{500, 1000\}$. Overall, our method performs remarkably better in all cases. More specifically, {\tt BGCWM} performs ideally in Scenarios 1 and 3, where the median number of misclassified significant/non-significant variables is either 0 or 1. Recall that both scenarios involve uncorrelated covariates, and typically it is  easier to distinguish relevant from irrelevant variables in such cases. In contrast, Scenario 2 is more challenging: the median number of misidentified variables is 4 (for $n=500$) and 2 (for $n=1000$). Here, covariates are correlated within clusters but indistinguishable across clusters (homogeneous), which complicates detection. Finally, in Scenario 4, the median number of misidentified covariates is 2 (for $n=500$) and 1 (for $n=1000$). As in Scenario 2, covariates are correlated within clusters; however, in this case, they are heterogeneous across clusters, making detection easier than in Scenario 2.   

\subsection{Simulations from mixture-of-experts model}\label{sec:moe_sim}

Here we seek to evaluate the robustness of our method for cases where the true data-generating process is not that of a cluster-weighted model. Namely,  we consider a scenario similar to Scenario 1  of Section \ref{sec:cwm_sim} (with uncorrelated homogeneous covariates), but instead of assigning each observation to a cluster with  fixed prior probability (as implied by the assumptions of the cluster weighted model), we assume that the mixing proportions depend on the covariates (as implied by the assumptions of the mixture of experts model).

Specifically, we assume that the proportions in Eq. \eqref{eq:z_prior} are given by 
\[
\pi_{ik}  ~ \propto ~  e^{\dot{\alpha}_k+\xxti\dot{\boldsymbol\beta}}_k,
\]
for $k = 1,\ldots,K$ and $i = 1,\ldots,n$. The coefficients $\dot{\alpha}_k$ and $\dot{\boldsymbol\beta}_k$ are generated using the same mechanism with $\alpha_k$ and $\boldsymbol{\beta}_k$ as in Section \ref{sec:cwm_sim}. Following the suggestion of an anonymous reviewer, the expert parameters $\alpha_k$ have been generated by a $t_3$ distribution, that is, a Student's $t$ distribution with three degrees of freedom. Finally, we considered that $n\in\{250, 500, 1000\}$ (sample size), $K\in{2, 3, 4}$ (number of clusters) and $p = 9$ covariates. For each distinct combination of $K$ and $n$, 30 replicated datasets have been generated. 

\begin{figure}[p]
    \centering
    \includegraphics[scale = 0.52]{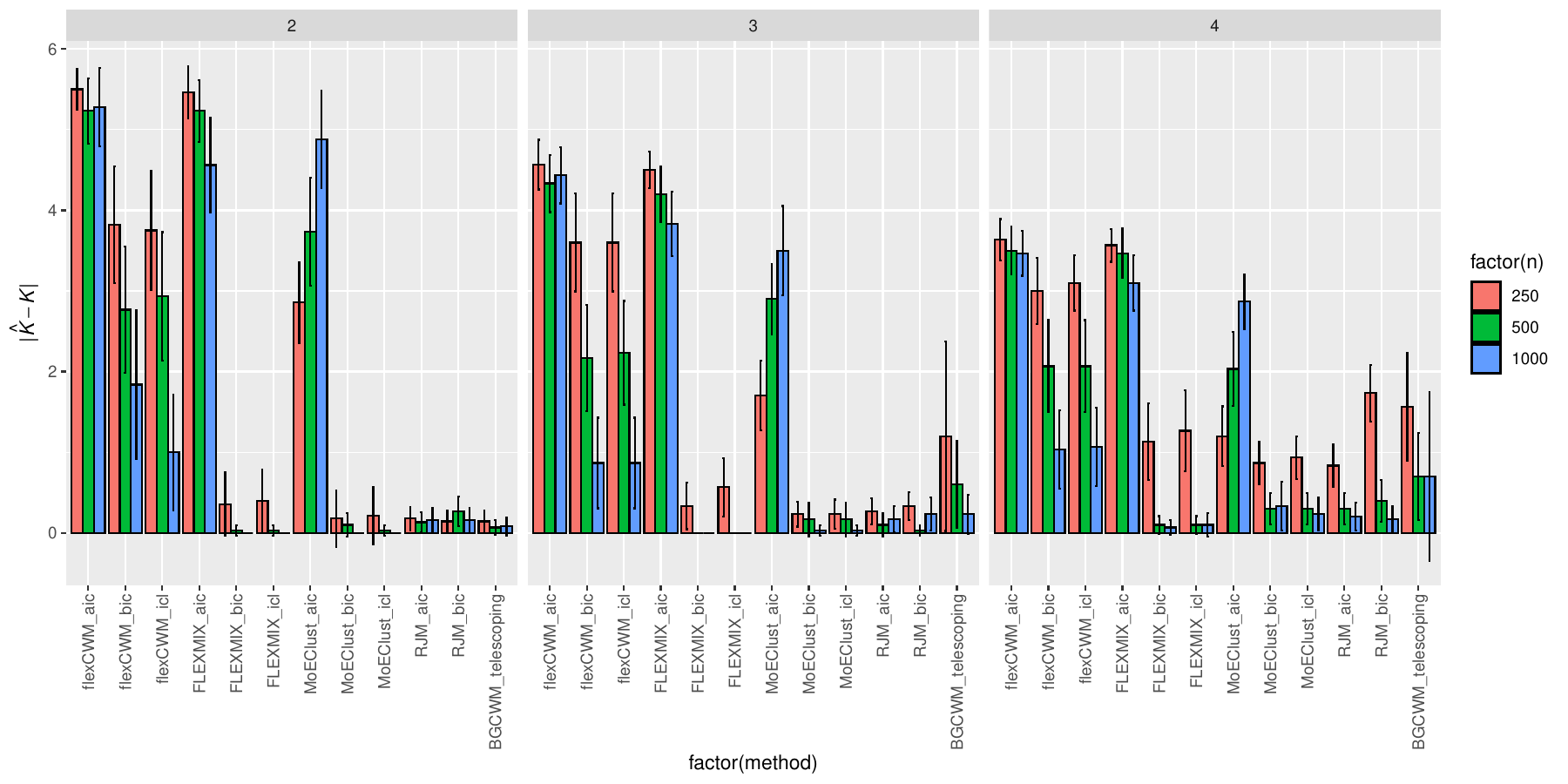}
    \caption{Mean absolute error of estimating the number of clusters  per value of $K\in\{2, 3, 4\}$ for the synthetic datasets of Section \ref{sec:moe_sim}. The error bars display (asymptotic) $95\%$ confidence intervals of the mean.}
    \label{fig:mae-moe}
\end{figure}

\begin{figure}[p]
    \centering
     \includegraphics[scale = 0.55]{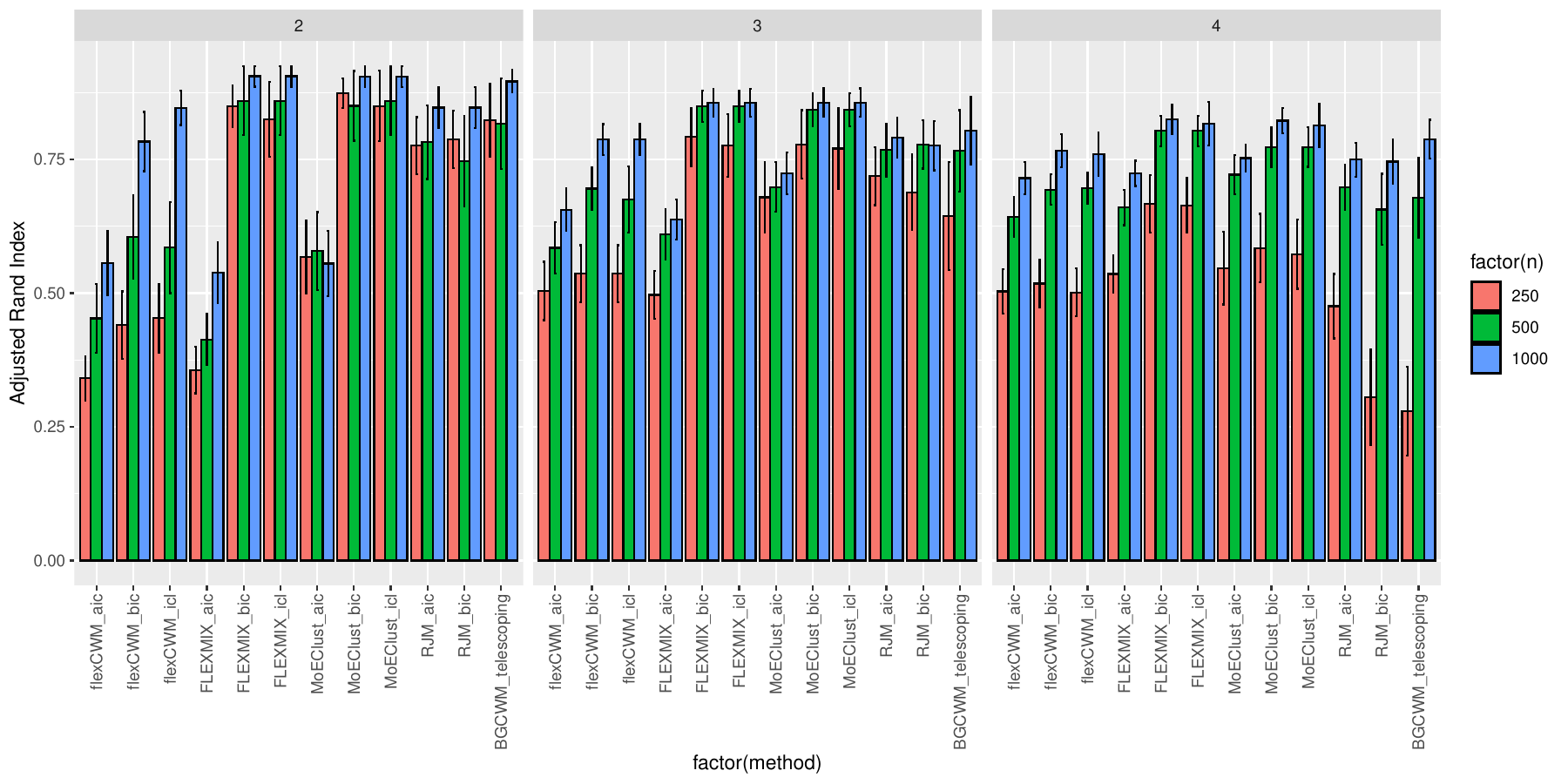}
     \caption{Mean adjusted Rand index  value of $K\in\{2, 3, 4\}$ for the synthetic datasets of Section \ref{sec:moe_sim}. The error bars display (asymptotic) $95\%$ confidence intervals of the mean.}
    \label{fig:ari-moe}
\end{figure}

The results are displayed in Figures \ref{fig:mae-moe} and \ref{fig:ari-moe}, in terms of Mean Absolute Error for selecting the number of clusters and Adjusted Rand Index with respect to the true cluster assignments, respectively. We conclude that when $K < 4$ and $n > 250 $ the proposed method can accurately detect the number of clusters and the clustering structure, resulting in a comparable performance to the {\tt MoEClust} and {\tt flexmix} implementations (which start from a correct assumption about the  ``true'' model structure). This even holds true for smaller values of sample size ($n=250$) and small number of clusters $(K = 2)$. However, in cases of small sample size ($n = 250$) and larger number of clusters ($K = 4$) the proposed method converges to larger values of number of clusters than the ones selected by {\tt flexmix} and {\tt MoEClust}  (which both use concomitant variables). 

\section{Application to TCGA data}\label{sec:real}

\begin{figure}[h]
    \centering
    \includegraphics[scale = 0.6]{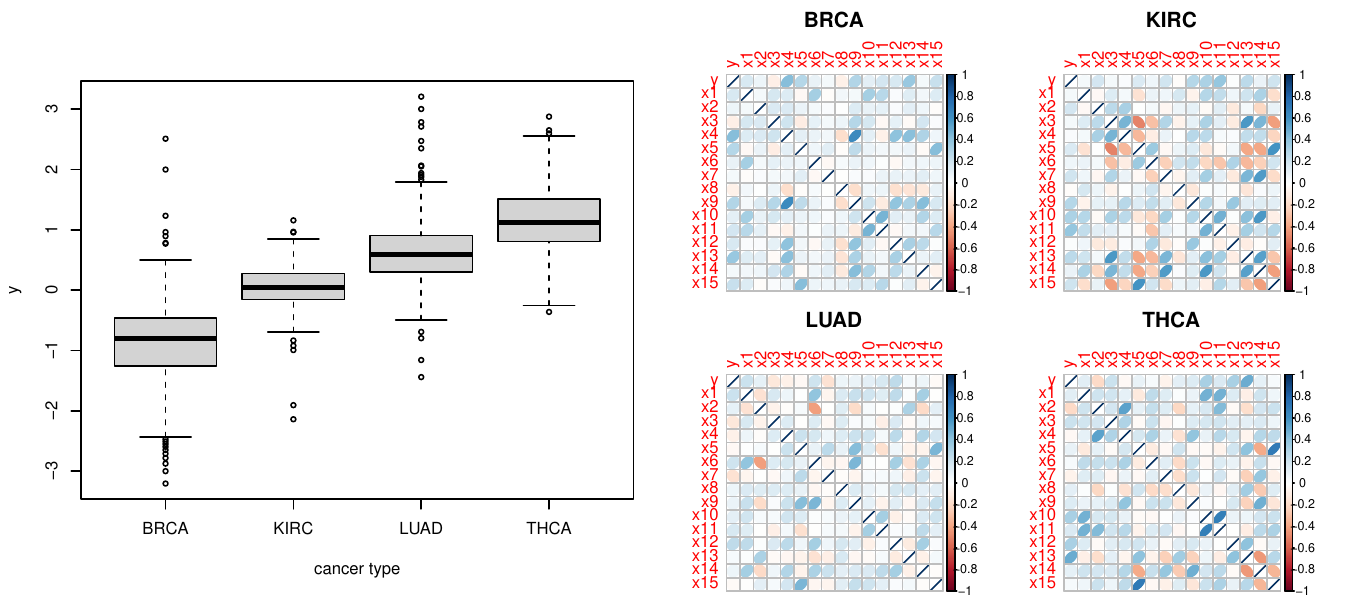}
    \caption{TCGA data: boxplots of the response gene (left) and pairwise Pearson correlations of the response gene and of the genes used as covariates (right) per cancer type.}
    \label{fig:real_data}
\end{figure}

In this section we illustrate the proposed BGCWM model for clustering  genomic data.  For this purpose we consider a dataset from The Cancer Genome Atlas (TCGA, \url{https://cancergenome.nih.gov}), consisting  of gene expression levels  from four cancer types: breast (BRCA), kidney renal clear cell (KIRC), lung adenocarcinoma (LUAD) and thyroid (THCA). The sample size is equal to $n=1485$ and the ground truth classification of the data into four cancer types corresponds to class frequencies equal to 608 (BRCA), 303 (KIRC), 288 (LUAD) and  286 (THCA). Similarly to the analysis in \cite{perrakis2023regularized}, here we consider the expression of gene GALNT12 (polypeptide N-acetylgalactosaminyltransferase 12; gene ID 79695)  as the response variable and  select at random $p=15$ other genes as covariates.  All variables are standardized, having zero mean and unit variance. Boxplots of the response variable are shown in Figure \ref{fig:real_data}, along with the corresponding correlation plots among the covariates and the response per cancer type. 

In what follows, we consider three inferential tasks: cluster  recovery (Section \ref{sec:cluster}), post-hoc variable evaluation (Section \ref{sec:variable_sel}) and predictive inference (Section \ref{sec:pred}). More detailed insights for the influential genes are given in Appendix D.1. Finally, Appendix D.2 discusses model diagnostics. 

\subsection{Clustering inference}\label{sec:cluster}

\begin{figure}[h]
    \centering
    \includegraphics[scale = 0.45]{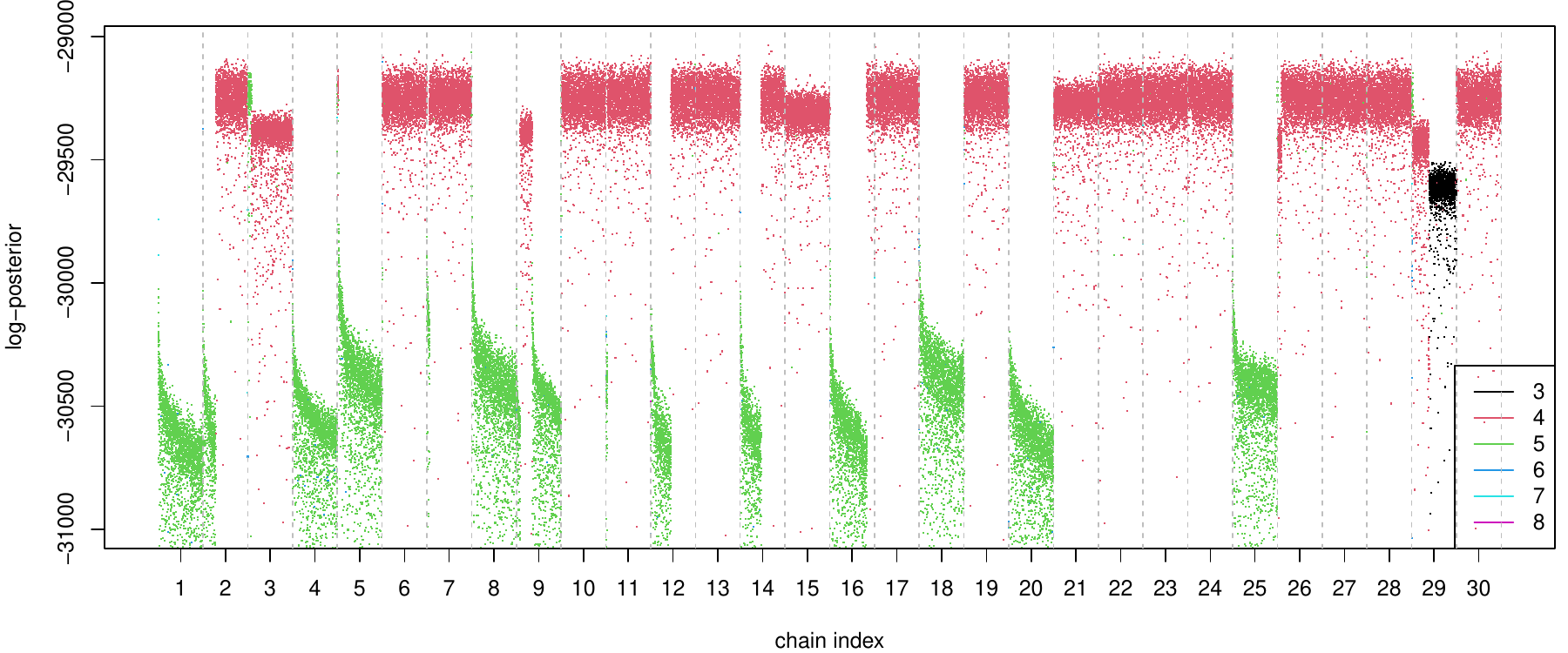}
    \caption{TCGA dataset: sampled values of the logarithm of posterior density (up to a multiplicative constant) across 30 telescoping-sampler runs. Colours correspond to different numbers of clusters (see legend). Each segment arises from a sample of 200000 MCMC draws (thinned at every 100th iteration) following a warm-up period of 50000 iterations.}
    \label{fig:real_multiple_chains}
\end{figure}

We implemented 30 runs of the telescoping sampler, each one consisting of 250000 iterations. Figure \ref{fig:real_multiple_chains} displays the sampled values of the logarithm of the posterior density for each run. It is evident that in several instances the  sampler is stuck within minor modes of the posterior distribution. In particular 12 out of 30 chains (namely, the chains 1, 3, 4, 5, 8, 9, 15, 18, 20, 21, 25 and 29) remain trapped in minor modes. Also, some chains require many MCMC iterations to discover the main mode; an example of this behavior is shown in the draws of chains 12 and 14 which both switch from a model with $K_+=5$ clusters (green points) to a model with $K_+=4$ clusters (red points) after a large number of MCMC iterations (note that the trace corresponds to thinned samples of 200000 iterations, following a warm-up period of 50000 iterations). We observed that this behavior is characterized by the presence of excessive clusters having few allocated observations.  The telescoping sampler has to, therefore, cross regions of small posterior probability in order to ``eliminate'' the excessive clusters and then  discover higher posterior density areas.

\begin{figure}[ht]
    \centering
    \begin{tabular}{cc}
    \includegraphics[scale = 0.4]{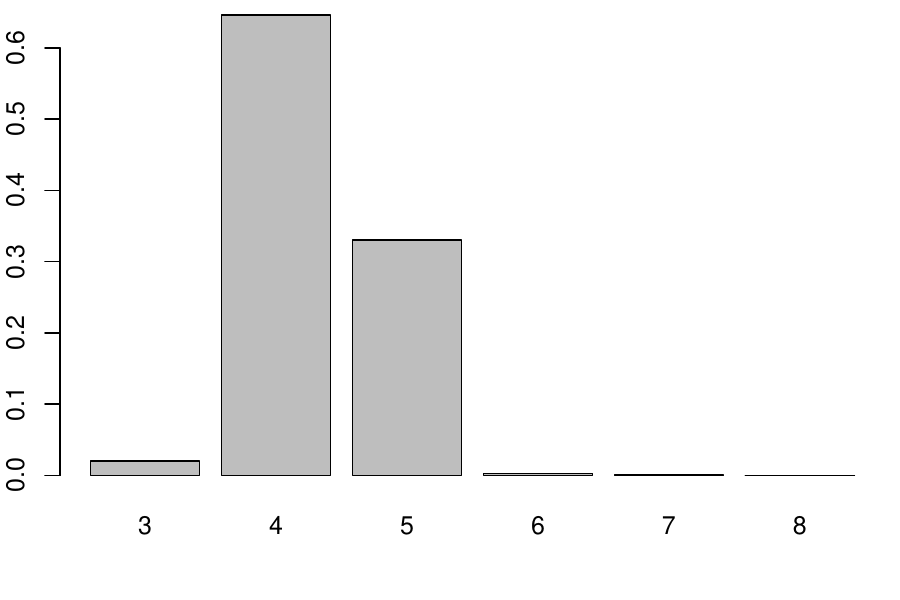}&    \includegraphics[scale = 0.4]{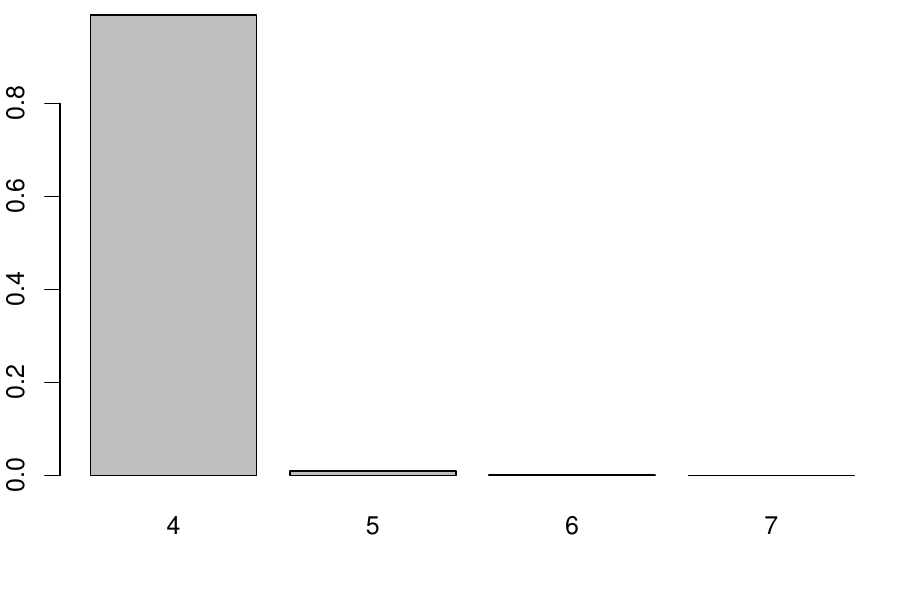}\\
    (a) & (b)
    \end{tabular}
    \caption{TCGA dataset: Posterior distribution of the number of clusters $K_+$ based on (a) 30 runs of the telescoping sampler and (b) the runs that converged to the main mode.}
    \label{fig:real_K_posterior}
\end{figure}

Figure \ref{fig:real_K_posterior}(a) displays the estimated posterior distribution of the  number of clusters $K_+$ arising from the 30 runs of the telescoping sampler. Figure \ref{fig:real_K_posterior}(b) displays the estimated posterior distribution of the number of clusters $K_+$, after keeping all runs that converged to the main mode of the posterior distribution. In both cases, the mode corresponds to $K_+ = 4$, which is consistent with the ``true'' number of groups (cancer types).  

Since the posterior distribution exhibits genuine multimodality, it is not meaningful to draw inferences based on the entire MCMC sample. Instead, we focused on distinct modes of the posterior surface  \citep{grun2009dealing, papastamoulis2010artificial}, after discarding all runs that failed to converge to the main mode of the posterior distribution. The retained MCMC draws were post-processed via the ECR algorithm in order to undo label switching, conditional on the most probable value ($K_+=4$). We then estimated the single best clustering of the observations; the resulting adjusted Rand index (with respect to the ground-truth classification) is equal to $0.662$. The corresponding confusion matrix is shown in Table \ref{tab:confusion_matrix}; note that the labeling is such that the rows $\{1, 2, 3, 4\}$ align with the columns $\{\mathrm{BRCA}, \mathrm{KIRC}, \mathrm{LUAD}, \mathrm{THCA}\}$ as much as possible. Following the suggestion of an anonymous reviewer, Table \ref{tab:confusion_matrix}  also reports the corresponding confusion matrix conditional on  a $K_+ = 5$ class partition, arising from the output of the telescoping sampler.

\begin{table}[ht]
\centering
\begin{tabular}{l|rrrr|rrrrr}
\hline
 & \multicolumn{4}{c|}{\textbf{$K_+=4$ clusters}} 
 & \multicolumn{5}{c}{\textbf{$K_+=5$ clusters}} \\
\cline{2-10}
 & 1 & 2 & 3 & 4 & 1 & 2 & 3 & 4 & 5 \\
\hline
BRCA & 533 &  30 &  45 &   0 & 543 & 23 & 42 &  0 &  0\\
KIRC &   6 & 284 &  13 &   0 & 10 &280  &10 &  1  & 2\\
LUAD &  14 & 125 & 147 &   2 & 17 & 97 &170 &  4  & 0\\
THCA &   1 &  10 &  12 & 263 &  1 &  9 &  9 &267  & 0\\
\hline
\textbf{Adjusted Rand Index} 
 & \multicolumn{4}{c|}{0.662} 
 & \multicolumn{5}{c}{0.692} \\
\hline
\end{tabular}
\caption{TCGA dataset: confusion matrices and Adjusted Rand Indices comparing ground-truth cancer types with estimated clusterings for $K_+=4$ and $K_+=5$, arising from the output of the telescoping sampler.}
\label{tab:confusion_matrix}
\end{table}

\subsection{Post-hoc variable evaluation}\label{sec:variable_sel}
Next, for each cluster, we evaluated which variables  influence the response after computing $90\%$ simultaneous credible intervals within each cluster and picking the significant ones according to Equation \eqref{eq:vs}. 
Our results indicate nine genes which have a significant impact on the response, GALNT12, across clusters. The influential genes are summarized in Table \ref{tab:genes}; as seen, the effects to the response gene differ among the four clusters. The posterior distributions of the regression coefficients can be found in Appendix D (Figure D.1; see also Table D.1). Classification residuals confirm that the model provides an adequate fit (see Appendix D.2 and Figure D.2). 

\addtolength{\tabcolsep}{-0.4em}
\begin{table}[H]
\centering
\begin{tabular}{c|rrrrrrrrr}
  \hline
  Cluster & \multicolumn{8}{c} {Influential genes}\\ 
  \hline
1 & \small{WNK4} & \small{SLC17A7} & \small{LRRC25} & \small{} & \small{} & \small{PI4K2B} & \small{CLDN10} & \small{} & \small{P2RY10} \\
2 & \small{} & \small{} & \small{} & \small{SLFNL1} & \small{} & \small{PI4K2B} & \small{} & \small{} & \small{} \\ 
3 & \small{} & \small{} & \small{} & \small{SLFNL1} & \small{} & \small{} & \small{} & \small{} & \small{}  \\ 
4 & \small{} & \small{SLC17A7} & \small{} & \small{} & \small{ARHGEF15} & \small{} & \small{CLDN10} & \small{FAM54B} & \small{}  \\ 
   \hline
\end{tabular}
\caption{TCGA dataset: influential genes per cluster (with $K_+= 4$ clusters).}
\label{tab:genes}
\end{table}

\subsection{Predictive inference}\label{sec:pred}
We will conclude the presentation of this  application by evaluating the predictive accuracy of our model, taking also into account the performance of other methods (including machine-learning methods as described below). For this aspect of the analysis, the TCGA dataset ($n = 1485$ observations in total) was randomly split into training and test subsets of various sizes (shown in Table \ref{tab:pred_scen}) and we considered only the telescoping sampler  ({\tt BGCWM}), {\tt FLEXMIX}   and {\tt MoEClust}\footnote{We note that {\tt RJM} and {\tt flexCWM} do not have a built-in function for prediction.}. As previously, for {\tt FLEXMIX} and {\tt MoEClust} we  considered information criteria (AIC, BIC, ICL). In both approaches we included concomitant variables\footnote{The corresponding implementations without concomitant variables gave inferior results.}. The telescoping sampler ({\tt BGCWM}) ran for 60000 iterations and the first 10000 iterations were discarded as burn-in period. The retained  MCMC sample was further thinned by keeping every 10th iteration. The posterior predictive distribution was estimated from the thinned MCMC sample of 5000 iterations. Predictions were then performed by estimating the mean of the posterior predictive distribution for each input value of covariates.  

In addition to the preceding approaches, we also explored non-parametric and machine learning methods. Specifically, generalized additive models \citep[implemented in the  \textsf{R}  package {\tt mgcv};][]{gam1, gam2}, Bayesian Additive Regression Trees \citep[\textsf{R} package {\tt BART};][]{BART}, extreme gradient boosting  \citep[\textsf{R} package {\tt XGBoost};][]{xgboost}, local approximate Gaussian processes  \citep[\textsf{R} package {\tt laGP};][]{laGP}, random forests \citep[\textsf{R} package {\tt randomForest};][]{rf} and feed-forward neural networks \citep[\textsf{R} package {\tt nnet};][]{nnet}. For {\tt BART}, {\tt mgcv} and {\tt randomForest} we employed the default parameter settings. We used cross-validation to optimally select parameters in the following methods:  {\tt laGP} (neighborhood size), {\tt nnet} (all hyper-parameters) and {\tt XGBoost} (optimal number of boosting rounds).

\begin{table}[p]
\centering
\begin{tabularx}{0.5\textwidth}{lccX}
    \toprule
    Scenario\ \ \ \ \ \  & $n_{\mathrm{train}}$\ \ \ \ \ \  & $n_{\mathrm{test}}$\ \ \ \ \ \  & \# datasets \\
    \midrule
    A& 1085 & 400 & 50 \\
    B & 1185 & 300 & 50  \\
    C & 1285 & 200 & 50 \\
    D & 1385 & 100 & 50 \\
    \bottomrule
\end{tabularx}
\caption{Sizes of train and test datasets of Section \ref{sec:pred}. }
\label{tab:pred_scen}
\end{table}
\begin{figure}[p]
    \centering
    \begin{tabular}{c}
         \includegraphics[scale = 0.57]{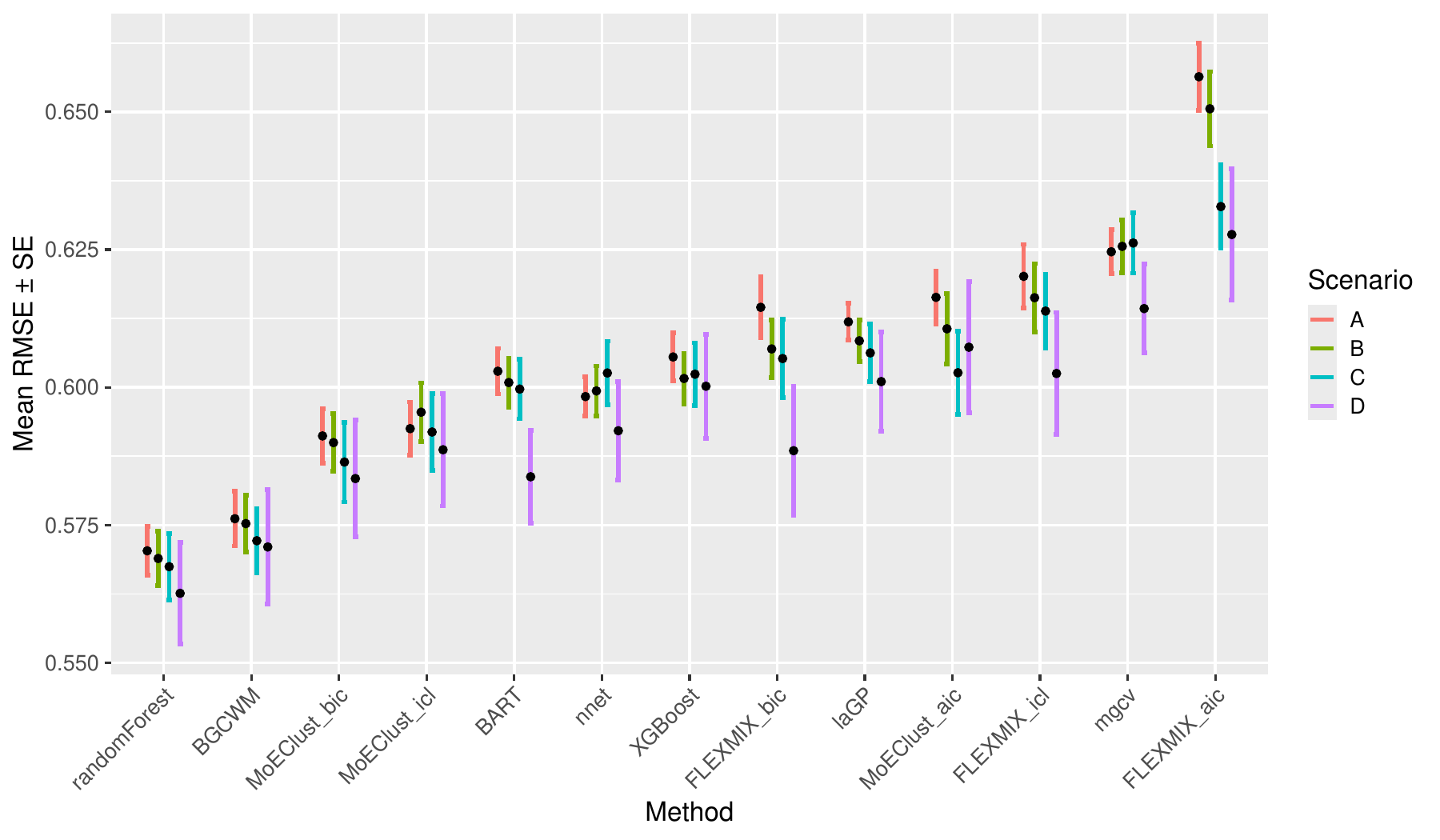}    
         \end{tabular}
     \caption{TCGA dataset: The dots denote the mean Root Mean Square Error (RMSE) computed over 50 randomly generated train–test splits, while the error bars represent the corresponding standard errors. Scenarios A, B, C, and D correspond to different training and test dataset sizes (see Table \ref{tab:pred_scen} for details). }
    \label{fig:pred}
\end{figure}

We repeated the procedure for a total of 50 random splits of the data (per scenario). The results are summarized in Figure \ref{fig:pred} in terms of Root Mean Square Error (RMSE). The methods are sorted in increasing order according to their overall average error. Our method ({\tt BGCWM}) compares favorably with the competing approaches, achieving the second-highest ranking, just behind {\tt randomForest}. We note however that the small predictive trade-off is outweighed by the interpretability and clustering ability that the proposed model-based method provides.


\section{Discussion}
\label{sec:discussion}

In this study we developed a fully Bayesian treatment of cluster weighted models with Gaussian covariates. The introduced model exploits shrinkage prior distributions both in the coefficients of the linear predictor within each cluster via the Bayesian lasso prior \citep{park_casella}, as well as the covariance of the random covariates via the Bayesian graphical lasso \citep{Wang2012}. We note that the proposed model is modular in nature, in the sense that different shrinkage methods can be applied as future extentions and variations; for instance, a combination of horseshoe priors for the regression coefficients \citep{carvalho_etal_2010} and the covariances of the predictors \citep{li_etal_2019}.  

The flexibility provided by generalized mixtures of finite mixtures allows us to treat the number of mixture components as random and draw fully Bayesian inference using the telescoping sampler of \cite{fruhwirth2021generalized}. We also discussed alternative approaches such as the overfitting mixture models \citep{rousseau2011asymptotic} as well as estimating separate models for each distinct value of the number of components and selecting one according to information criteria. 


Naturally, our model is invariant to label switching, therefore posterior inference conditional on a given value of number of clusters is meaningful only after the MCMC output is suitably post-processed. The Equivalence Classes Representatives (ECR) algorithm \citep{papastamoulis2010artificial, papastamoulis2016} was used for this purpose. After this step, we followed the approach of \cite{Papastamoulis2022} in order to select relevant variables within each cluster, by computing simultaneous credible intervals. One could fully address the issue of variable selection, for example by exploiting stochastic search variable selection \citep{george1995stochastic, ntzoufras_e_tal_2000, dellaportas_etal_02} approaches. We plan to explore this issue in the future. 

In its current form the proposed model is designed for continuous covariates. Therefore, an interesting further direction of future research relates to mixed sets of data types. Obviously, it would be quite challenging to efficiently describe covariance patterns between discrete and continuous data in such settings. In addition, the generalization to other types of response variables, e.g., categorical or count responses as considered in  \cite{ingrassia2015erratum, punzo2015parsimonious} and/or multivariate responses, as considered in \cite{murphy_murphy2020}, is  another interesting future direction worth of exploration. 

Finally, one could also consider the question of improving the mixing of the MCMC sampling. For instance, by considering parallel tempering schemes such as Metropolis-coupled MCMC.

\section*{Data availability}
The genomic dataset used in Section \ref{sec:real} is extracted from  (TCGA, \url{https://cancergenome.nih.gov}) and it is available online at  \url{https://github.com/mqbssppe/BGCWM}.

\section*{Acknowledgments}
We are grateful to the two anonymous reviewers for their insightful comments and suggestions, which helped us improve both the presentation and the findings of this work.

\section*{Ethics declarations}
\subsection*{Conflict of Interest}

The authors declare no competing interests.

\section*{Author Contribution}

The first author was responsible for conceptualization, methodology, investigation, software development, simulation studies, data analysis and visualization, writing, review, and editing. The second author was responsible for conceptualization, methodology, investigation, writing, review, and editing.

\section*{Supplementary material}

\begin{description}

\item[{\tt appendix.pdf}: ] The online Appendix includes the following
\begin{itemize}
    \item[Appendix A:] presents the steps of the telescoping sampler in pseudocode.
    \item[Appendix B:] provides details on the MCMC implementation and the initialization of the algorithms.
    \item[Appendix C:]  reports more detailed results from the simulation study.
    \item[Appendix D:]  additional results for the TCGA dataset, that is,
    \begin{itemize}
        \item[Appendix D.1:] presents additional results regarding influential genes.
        \item[Appendix D.2:] asses model fit using residual diagnostics. 
    \end{itemize}
    \item[Appendix E:]  offers insights into the computational cost of the telescoping sampler in terms of CPU time. 
\item[Appendix F:]  describes the underlying statistical models implemented in each software package ({\tt FLEXMIX}, {\tt MoEClust}, {\tt flexCWM}, {\tt RJM}) considered in the comparisons of Section \ref{sec:sim}.     
\end{itemize}

\item[Source code:] The \textsf{R} source code used in this manuscript is available online at \url{https://github.com/mqbssppe/BGCWM}, together with scripts that exemplify its usage in simulated as well as our real datasets.  The authors are also planning to release the code in the form of an \textsf{R} package.

\end{description}


\bibliographystyle{chicago}
\bibliography{biblio_R3}

\end{document}